\documentclass[9pt,twocolumn,twoside,lineno]{article}

\usepackage{fullpage}
\usepackage{graphicx}
\usepackage{amsmath}
\usepackage{amssymb}
\usepackage{mathrsfs}
\usepackage{float}
\usepackage[utf8]{inputenc}
\usepackage{listings}
\usepackage{color}
\usepackage{amsfonts}
\usepackage{array}
\usepackage{natbib}
\usepackage{booktabs}
\usepackage{titlesec}
\usepackage[margin=1.0in]{geometry}
\usepackage[switch]{lineno}

\let\originaleqref\eqref
\renewcommand{\eqref}{Eq.~\originaleqref}

\setcitestyle{numbers}

\titlespacing\section{0pt}{12pt plus 4pt minus 2pt}{0pt plus 2pt minus 2pt}
\title{A general constitutive model for dense, fine particle suspensions validated in many geometries}

\author{A. Baumgarten$^1$, K. Kamrin$^1$\\
	\small{$^1$ Massachusetts Institute of Technology, Cambridge, MA 02139}}

\date{This manuscript was compiled on \today}

\begin{document}
	
\twocolumn[
\begin{@twocolumnfalse}
	\maketitle
	\begin{abstract}
		Fine particle suspensions (such as cornstarch mixed with water) exhibit dramatic changes in viscosity when sheared, producing fascinating behaviors that captivate children and rheologists alike. Recent examination of these mixtures in simple flow geometries suggests inter-granular repulsion is central to this effect --- for mixtures at rest or shearing slowly, repulsion prevents frictional contacts from forming between particles, whereas, when sheared more forcefully, granular stresses overcome the repulsion allowing particles to interact frictionally and form microscopic structures that resist flow. Previous constitutive studies of these mixtures have focused on particular cases, typically limited to two-dimensional, steady, simple shearing flows. In this work, we introduce a predictive and general, three-dimensional continuum model for this material, using mixture theory to couple the fluid and particle phases. Playing a central role in the model, we introduce a micro-structural state variable, whose evolution is deduced from small-scale physical arguments and checked with existing data. Our space- and time-dependent model is implemented numerically in a variety of unsteady, non-uniform flow configurations where it is shown to accurately capture a variety of key behaviors: (i) the continuous shear thickening (CST) and discontinuous shear thickening (DST) behavior observed in steady flows, (ii) the time-dependent propagation of `shear jamming fronts', (iii) the time-dependent propagation of `impact activated jamming fronts', and (iv) the non-Newtonian, `running on oobleck' effect wherein fast locomotors stay afloat while slow ones sink.
	\end{abstract}
\end{@twocolumnfalse}
]

\section{Introduction}
The behavior of granular materials suspended in fluid media has been a major topic of study for over a century. These types of mixtures are present in many industrial, geotechnical, and biological engineering problems, spanning length scales from tens of meters to millimeters. Of particular interest in this work is the behavior of chemically stable, hard, frictional particles suspended in viscous fluids as found in industrial processes and studied in soil mechanics.

Though thoroughly studied and classified, a unifying constitutive model relating mixture stresses, strains, and strain rates across all material types and flow regimes remains elusive (see \cite{stickel2005}). One challenge for such a model is capturing the non-Newtonian, shear thickening behavior that is often observed in mixtures like water-cornstarch and water-poly(methyl methacrylate) when the mean particle diameter, $d$, {{\color{blue}} is less than $\sim$10 $\mu$m}. At low volume fractions, the apparent viscosity of such mixtures grows steadily with increasing shearing rate (CST); however, at high volume fractions, the apparent viscosity of these mixtures can jump several orders of magnitude with very little change in measured shearing rate (DST).

Recent theoretical work in \cite{wyart2014}, experimental observations in \cite{guy2015}, \cite{clavaud2017}, \cite{fall2012}, and \cite{brown2012}, and simulations reported in \cite{seto2013}, \cite{mari2014}, \cite{mari2015discontinuous}, and \cite{singh2018} have shown that the shear thickening behavior observed in these mixtures is a direct result of inter-granular repulsion and its effect on the dilation behavior of these mixtures. In the presence of relatively small applied stresses, the particles in the mixture will interact through lubrication forces in the suspending medium and behave like a granular material with low internal friction. In the presence of relatively large applied stresses, the particles in the mixture will be forced into frictional contact, drastically increasing the apparent internal friction coefficient. In dense suspensions, this frictional transition can cause the granular skeleton to dilate.

Prior work modeling general fluid-sediment mixtures in \cite{baumgarten2018} has focused on hard, frictional, non-Brownian, non-repulsive particle suspensions ($d\gtrsim100$ $\mu$m for common engineering mixtures). The modeling framework proposed in that work combines the empirical relations presented in \cite{stickel2005}, \cite{boyer2011} and \cite{amarsid2017} with the two-phase mixture theories developed in \cite{drumheller2000} and \cite{jackson2000} into a single constitutive theory. In this work, we build upon this framework to produce a more general model for granular suspensions that accounts for the effect of inter-granular repulsion on mixtures with mean grain diameters $d \lesssim 100$ $\mu$m. Using a custom numerical scheme, we validate this model in transient and inhomogeneous flows, in both two-dimensional and three-dimensional geometries. The model is shown to replicate the unusual landmark features of these shear-thickening suspensions such as the ability to run across them but not walk.

\section{Model} \label{sec:model}
\subsection*{General Theory} 
In continuum modeling of fluid-{{\color{blue}}particle} mixtures, we consider the materials that constitute the mixture independently. The individual solid particles (or grains) are {homogenized} into a single continuum body called the \textit{granular phase} which has a density $\bar{\rho}_s = \phi \rho_s$ and velocity components ${v_s}_i$. Here $\phi$ defines the volume fraction of the {{\color{blue}} granular phase }and $\rho_s$ defines the density of each individual grain. Similarly, the fluid which fills the space between the grains is {homogenized} into a single continuum body called the \textit{fluid phase} with density $\bar{\rho}_f = (1-\phi)\rho_f$ and velocity components ${v_f}_i$. Here $\rho_f$ is called the \textit{true} density of the fluid. A thorough theoretical description of this process of {homogenization} can be found in \cite{baumgarten2018} and \cite{jackson2000}.

From the basic laws of mass conservation, momentum and energy balance, and entropy imbalance, it is possible to define a complete set of governing equations for the evolution of density, velocity, and stress within each phase. These equations account for (i) momentum exchange between the phases through a buoyant force and a Darcy-like inter-phase drag, (ii) dilation and contraction of the granular phase through a specialized elastic-plastic constitutive model, and (iii) separation of mixture stresses into components from the granular skeleton, $\tilde{\sigma}_{ij}$, and components from the pore fluid, ${\tau_f}_{ij}$ and $p_f$. These equations are combined to form the model proposed in \cite{baumgarten2018}. This model is shown to be accurate in a wide range of flow regimes and geometries; however, it does not account for the effects of inter-granular repulsion and corresponding structure evolution. In this work, we show that reformulating two previously constant parameters from the model shown in \cite{baumgarten2018}, $a$ and $\phi_m$, is sufficient to capture the physics of particle repulsion and accurately model steady CST and DST as well as several transient and dynamic behaviors observed in the dense, fine grain suspensions we seek to model.

The full set of equations that define our proposed three-dimensional, time-dependent, two-phase model can be found in the Supplemental Information; however, several key features of this model are easily illustrated by considering its behavior in steady two-dimensional (or quasi-two-dimensional) shearing flows. This behavior is described by the relationship between the steady mixture shearing rate $\dot{\gamma}$, the mixture shear stress $\tau$, and the granular phase pressure $\tilde{p}$ . For flows dominated by viscous effects (e.g.\ water-cornstarch; $(\rho_s d^2 \dot{\gamma})/\eta_0 \ll 1$), {{\color{blue}} that steady state relationship is identical to the model proposed in \cite{morris1999} and presented in \cite{boyer2011} and is expressed as follows},
\begin{equation} \label{eqn:effective_viscosity}
\frac{\tau}{\eta_0 \dot{\gamma}} = \eta_r =  1 + \frac{5}{2} \phi \bigg(\frac{\phi_m}{\phi_m - \phi}\bigg) + 2 \mu_c \bigg(\frac{a\phi}{\phi_m - \phi}\bigg)^2,
\end{equation}
\begin{equation} \label{eqn:normal_viscicity}
\frac{\tilde{p}}{\eta_0 \dot{\gamma}} = \eta_n = 2 \bigg(\frac{a \phi}{\phi_m - \phi}\bigg)^2
\end{equation}
with $\rho_s$ the density of the grains, $\eta_0$ the fluid phase viscosity, $\phi$ the granular volume fraction, and $\mu_c$ the flow resistance due to frictional granular contacts. $\phi_m$ and $a$ are dilation parameters and have a significant effect on the mixture stress and granular phase pressure.

The parameter $\phi_m$ limits the range of volume fractions where steady flow is possible (i.e.\ for $\phi \geq \phi_m$, shearing the material results in continuous growth of the shear stress, sometimes referred to as a \emph{jammed} state) and the parameter $a$ modifies the critical state volume fraction (see $\phi_{eq}$ in \cite{baumgarten2018} and Supplemental Information) and associated Reynolds' dilation. $\phi_m$ and $a$ are often treated as material constants; however, as proposed in \cite{wyart2014} and fully defined in \cite{singh2018}, $\phi_m$ and $a$ are better described as functions with $\phi_m$ bounded between $\phi_j$ and $\phi_c$ (with $\phi_j > \phi_c$) as follows,
\begin{equation} \label{eqn:phim_func}
\phi_m = \hat{\phi}_m(f) = \phi_j + (\phi_c - \phi_j)f,
\end{equation}
and $a$ bounded between $a_0$ and $a_\infty$ as follows,
\begin{equation} \label{eqn:a_func}
a = \hat{a}(f) = a_0 + (a_\infty - a_0)f,
\end{equation}
with $f$ a scalar measure of the anisotropy within the granular structure and the fraction of frictional granular interactions (bounded between 0 and 1). Conceptually, the material parameters $\phi_j$ and $\phi_c$ correspond to the range of volume fractions where steady shearing flow can occur. For $\phi < \phi_c$, steady flow can occur at \emph{all} shearing rates. For $\phi \geq \phi_j$, no steady flow is possible; the mixture is \emph{jammed}. For $\phi_c \leq \phi < \phi_j$, steady flow is only possible at \emph{some} shearing rates. $a_0$ and $a_\infty$ are fitting parameters.

Unlike in previous models where $f$ was a function of granular stress directly, we propose that $f$ evolves over time according to the following rule,
\begin{equation} \label{eqn:c_model}
\frac{\dot{f}}{K_0 \dot{\gamma}} = H\ (f_m - f) - Sf,
\end{equation}
with $K_0$ a dimensionless parameter, $H$ a dimensionless hardening rate, $S$ a dimensionless softening rate, and $f_m$ an upper bound on the steady value of $f$ given by $f_m = \hat{f}_m(\phi).$ Such an upper bound on $f$ was first proposed for low volume fractions in \cite{royer2016} and is an important feature in this model, especially as it relates to \emph{rheo-chaos} (see \cite{hermes2016} and \cite{grob2016}).

The first term in \eqref{eqn:c_model}, $H\ (f_m - f)$, describes the evolution of $f$ toward $f_m$ and, by \eqref{eqn:effective_viscosity} and \eqref{eqn:phim_func}, the associated increase in effective viscosity, $\eta_r$. Based on the work of \cite{mari2014} (see Supplemental Information), we propose the following form for $H$,
\begin{equation} \label{eqn:hardening}
H = \bigg(\frac{\bar{\tau}}{\tau^*}\bigg)^{3/2},
\end{equation}
%
%
with $\tau^*$ a repulsive stress scale with units of Pa (see $\tau_{\text{min}}$ in \cite{brown2014}, and $\sigma^*$ in \cite{singh2018}), $\bar{\tau} = (\mu_c + \mu_h) \tilde{p}$ the effective granular shear stress, and $\mu_h$ the flow resistance due to fluid-mediated granular interactions (see \cite{boyer2011}). This form of $H$ captures the strengthening of the granular phase as the granular stress forcing grains together overcomes inter-granular repulsion. 
The second term in \eqref{eqn:c_model}, $-Sf$, describes the evolution of $f$ toward 0 and the associated decrease in effective viscosity, $\eta_r$. We propose the following form for $S$,
\begin{equation} \label{eqn:softening}
S = 1 + \frac{\bar{\tau}}{\eta_B \dot{\gamma}} + \frac{\dot{\xi}_\epsilon}{\dot{\gamma}},
\end{equation}
with $\eta_B$ a measure of the resistance of the granular skeleton to buckling induced degradation with units of Pa$\cdot$s, and $\dot{\xi}_\epsilon$ a rate scale for Brownian or electrostatic repulsion with units s$^{-1}$. The three terms in $S$ reflect three proposed mechanisms which will break down granular structure. First, macroscopic shearing of the mixture will cause grains to slip past each other, altering their structure (see \cite{cates1998} for discussion of `plastic rearrangement' in jammed materials). 
{{\color{blue}}
Second, we postulate that force chains can undergo load-initiated reorganization ($\bar{\tau}/\eta_B \dot{\gamma}$) presumably through chain buckling, inducing structural degradation without macroscopic deformation. In systems of grains with large but finite elastic stiffness $G_p$, the small relative contact area $A_c/d^2$ of the particles means the buckling load $\tau_c\sim G_p A_c^2/d^4$ need not be large. The proposed form might then be seen as the product of a statistical measure of critical chains ($\sim \bar{\tau} /\tau_c$) and a time-scale of buckling degradation given by the properties of the mixture ($t_c \tau_c/\eta_0 = \hat{\varphi}(\rho G_p d^2/ \eta_0^2, A_c/d^2, \phi)$, i.e. $\eta_B\propto\tau_c t_c$).}
And lastly, Brownian or electrostatically driven diffusion will drive grains apart ($\dot{\xi}_\epsilon/\dot{\gamma}$); in flowing materials at room temperature, this contribution to $\dot{f}$ should be negligible.

\subsection*{Steady Behavior of Rheologically Stable Flows}
\begin{figure}[h]
	\centering
	\includegraphics[scale=0.6]{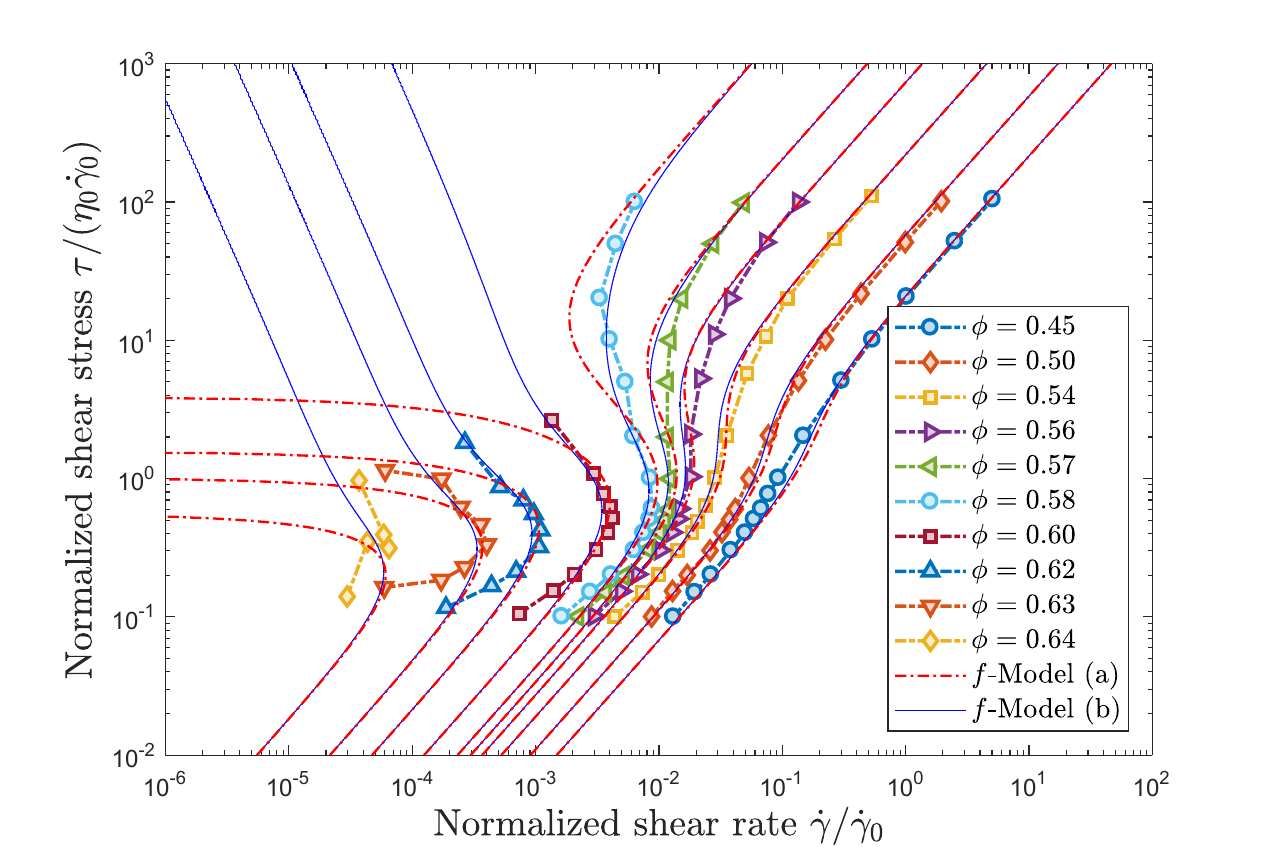}
	\caption{\footnotesize Comparison of steady shearing response of model described in section \ref{sec:model} to data reported in \cite{singh2018} for $\mu_c = 1$. Results are normalized by $\eta_0 \dot{\gamma}_0$ and associated shearing rate $\dot{\gamma}_0$ as reported in \cite{singh2018}. The simulated data are represented by the square, triangular, and circular markers (see legend). The response of our model is shown with red dashed lines for the fit parameters in table \ref{tab:singh}(a) and shown with blue solid lines for the fit parameters in table \ref{tab:singh}(b) at each of the packing fractions [0.45, 0.50, 0.54, 0.56, 0.57, 0.58, 0.60, 0.62, 0.63, 0.64].}
	\label{fig:singh}
\end{figure}

We begin analyzing the proposed model by considering the simulated three-dimensional shearing flows in \cite{singh2018}. These flows were composed of between 500 and 2000 stiff, spherical particles undergoing simple shear in a periodic domain. The discrete particle interactions were modeled using standard contact laws in conjunction with additional forces to account for lubrication and inter-granular repulsion. The steady state relationship between $\dot{\gamma}/\dot{\gamma}_0$ and $\tau/(\eta_0 \dot{\gamma}_0)$ (with $\eta_0 \dot{\gamma_0} \propto \tau^*$) reported in that work (see figure \ref{fig:singh}) contains several regimes where $\partial \dot{\gamma}/\partial \tau \leq 0$. These regimes can lead to unstable behavior under imposed shearing rates (see \cite{mari2015nonmonotonic}) and imposed shear stresses (see \cite{grob2016}). The latter instability manifests in rheologically chaotic behavior (i.e. transient inhomogeneities in the flow occur; \emph{rheo-chaos}). The lack of reported \emph{rheo-chaos} in \cite{singh2018} suggests that the behavior of larger systems of particles may differ at packing fractions near and above $\phi_c$.

We can, however, still examine the steady state response of the model in \eqref{eqn:c_model} for such rheologically stable shearing flows. To do this, we first define a functional form for $\hat{f}_m(\phi)$. The results in \cite{mari2014} are generated using the method shown in \cite{singh2018} and suggest that $f_m$ is insensitive to changes in $\phi$. We therefore let $f_m = 1$ for such mixtures. With $f_m$ determined, we calculate and fit the steady shear response of our model to the data in \cite{singh2018} by solving for $\dot{f} = 0$ at different volume fractions $\phi$ and shear rates $\dot{\gamma}/\dot{\gamma}_0$ (with $\tau/\eta_0 \dot{\gamma}_0$ determined from \eqref{eqn:effective_viscosity}) as shown in figure \ref{fig:singh}.
The relevant material parameters for these fits are provided in table \ref{tab:singh}. (See Supplemental Information for specific details about our model fitting procedure.)

\begin{table}[h]
	\small
        {\color{blue}}
		\centering
		\caption{\small Material parameters for curves in figure \ref{fig:singh}. $\mu_c=1$ for all fits. $\mu_h$ determined from $a$ and $\phi_m$ (see \cite{baumgarten2018}).}
		\label{tab:singh}
		\begin{tabular}{  c  c  c  c  c  c  c  } 
			Fit & $\phi_c$ & $\phi_j$ & $a_0$ & $a_\infty$ & $\tau^*/(\eta_0 \dot{\gamma}_0)$ & $\eta_0/\eta_B$\\
			\midrule
			(a) & 0.585 & 0.65 & 0.455 & 0.841 & 1.74 & 0\\
			(b) & 0.585 & 0.65 & 0.455 & 0.841 & 1.44 & 0.254$(\phi_j - \phi)$\\
			\bottomrule
		\end{tabular}
\end{table}

\subsection*{Pseudo-Steady Behavior of Rheologically Chaotic Flows}
We continue analyzing the proposed model by considering experimental analysis of repulsive grain mixtures. Although many authors have explored the phenomenology of these mixtures, complete characterization of their behavior is complicated by several factors which require additional modeling considerations beyond the ideal case discussed in the prior section. The scaling of $\tau^*$ with respect to mean grain diameter $d$ is often reported as $\tau^* \propto d^{-2} \text{ or }d^{-3}$ (see \cite{frith1996}, \cite{guy2015}, and \cite{brown2014}) and is observed to have vanishing effect on the response of common granular suspensions with $d \gtrsim 100$ $\mu$m; this puts practical limitations on the minimum system size (quantified by number of grains) that can be analyzed. Slip along boundaries (see \cite{fall2015} and \cite{peters2016}) and mixture breakdown are observed when applied shear stresses exceed approximately $10^4$ Pa, limiting the range of mixture responses that can be probed. Additionally, starches and other porous particles swell significantly in suspension, obfuscating the true volume fraction of the mixture (see \cite{hermes2016}).

Nevertheless, experimentation is the most direct way to examine the behavior of these particle suspensions and has yielded many important observations. {{\color{blue}} Among these is the \emph{rheo-chaos} reported in \cite{hermes2016}, \cite{dheane1993}, and \cite{boersma1991}}; these rheologically chaotic flows are characterized by large, rapid changes in the measured mixture shearing rate at constant applied stresses and measured relationships between $\dot{\gamma}$ and $\bar{\tau}$ that generally have $\partial \dot{\gamma}/\partial \bar{\tau} \geq 0$. More thorough study is necessary to understand the mechanics and transient effects of \emph{rheo-chaos}, but if we consider the long-time average behavior of the shearing flows reported in the literature to be representative of the average local behavior, we can examine the steady state response of the model in \eqref{eqn:c_model} for such rheologically chaotic flows.

We begin this analysis by defining a new expression for $\hat{f}_m(\phi)$ based on the experimental observations reported in \cite{hermes2016} and \cite{fall2015} as follows,
\begin{equation} \label{eqn:cmax}
\hat{f}_m(\phi) =
\begin{cases}
1,& \text{if } \phi \leq \phi_c\\
\frac{\phi - \phi_j}{\phi_c - \phi_j},& \text{if } \phi_c < \phi \leq \phi^*\\
\frac{\phi^* - \phi_j}{\phi_c - \phi_j},& \text{if } \phi > \phi^*,
\end{cases}
\end{equation}
with $\phi^* = \phi_j + (\phi_c - \phi_j)\Delta$
%
and $\Delta$ an implicit function of the compliance of the granular skeleton. By observation, we expect $\Delta \to 0$ for large systems of irregular grains (where \emph{rheo-chaos} is observed) and $\Delta \to 1$ for small systems of stiff, spherical grains (where \emph{rheo-chaos} is not observed). In the latter case, $f_m \to 1$ as in the idealized mixtures of the previous section.

Given this form of $\hat{f}_m(\phi)$ it is possible to determine a critical shearing rate associated with DST, $\dot{\gamma}_{\text{DST}}$, which is the limiting value of $\dot{\gamma}$ as $\bar{\tau}$ increases for the volume fractions in the range $\phi_c \leq \phi \leq \phi^*$ (see Supplemental Information).
Based on the distribution of apparent $\dot{\gamma}_{\text{DST}}$ values in the literature and the trends of $\dot{\gamma}_c$ reported in \cite{barnes1989}, we propose the following form of the structural resistance to buckling induced degradation $\eta_B$,
\begin{equation}\label{eqn:eta_B}
\eta_B = \hat{\eta}_B(\phi) = \bigg(\sum_{i=1}^{i_{\text{max}}} \varphi_i (\phi_j - \phi)^{\alpha_i} \bigg)^{-1}
\end{equation}
with $\varphi_i$ representing material parameters with units of (Pa$\cdot$s)$^{-1}$ and $\alpha_i$ representing dimensionless scaling factors. 

\begin{figure*}[!ht]
	\centering
	\includegraphics[scale=0.45]{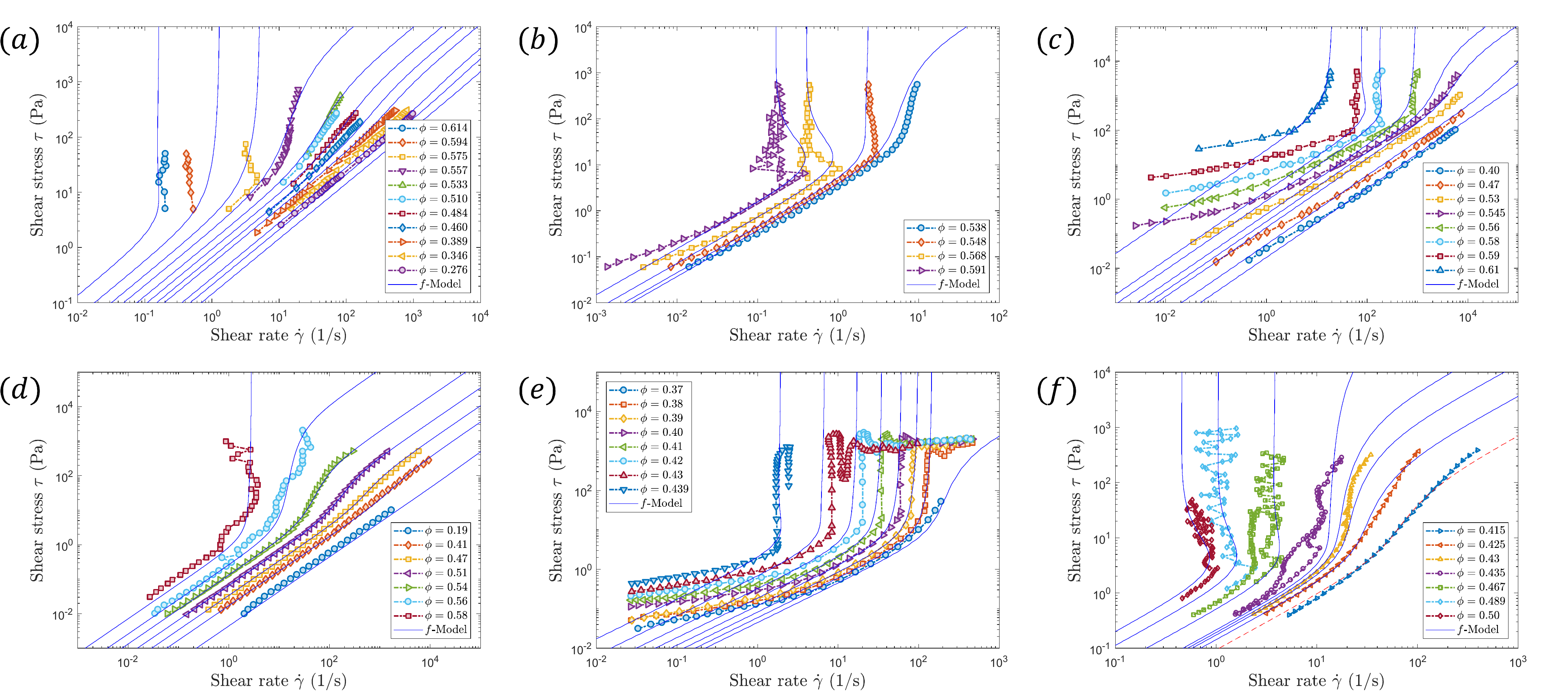}
	\caption{\footnotesize Comparison of steady shearing of model described in section \ref{sec:model} to experimental data reported in (a) \cite{dheane1993}, (b) \cite{frith1996}, (c) and (d) \cite{guy2015}, (e) \cite{fall2015}, and (f) \cite{hermes2016}. Experimental data are represented by colored markers at volume fractions noted in the associated legends. The response of our model is shown with the solid blue curves at the relevant volume fractions using the fit parameters in tables \ref{tab:experiment} and \ref{tab:experiment_etaB}. The dashed red curve in (f) corresponds to the response our model at a volume fraction of 0.39 using the fit parameters for the data in \cite{hermes2016}.} \label{fig:rheochaoticflows}
\end{figure*}

With these expressions determined, we calculate and fit the steady shear response of our model to experimental results reported in \cite{dheane1993}, \cite{frith1996}, \cite{guy2015}, \cite{fall2015}, and \cite{hermes2016} by solving for $\dot{f} = 0$ while satisfying the relationship in \eqref{eqn:effective_viscosity} at at the relevant volume fractions and shearing rates
as shown in figure \ref{fig:rheochaoticflows}. All data from the various experiments are seen to be well represented by the generic form proposed. The parameters for these steady shearing experiments can be found in tables \ref{tab:experiment} and \ref{tab:experiment_etaB}. {The wide range of $\tau^*$ values reported in table \ref{tab:experiment} is likely linked to the scaling of $\tau^* \propto d^{-2}$ or $d^{-3}$ and changes in surface chemistry between different granular materials {{\color{blue}} (see the effects of hydrogen bonding on `shear jamming' and DST in \cite{james2018})}. \cite{dheane1993}, \cite{frith1996}, and \cite{guy2015} performed experiments using poly(methyl methacrylate) beads with diameters $d$ between roughly 0.3$\mu$m and 4$\mu$m while \cite{fall2015} and \cite{hermes2016} performed experiments using cornstarch with reported grain diameters often between 5$\mu$m and 20$\mu$m.} (See Supplemental Information for specific details about fitting $\dot{\gamma}_{\text{DST}}$ and determining $\phi$ from \cite{hermes2016}.)

\begin{table}[h]
	\small
	\begin{center}
		\caption{\small Material parameters for curves in figure \ref{fig:rheochaoticflows}. $\mu_c = 0.3$ for \cite{dheane1993}, \cite{frith1996} and \cite{guy2015}. $\mu_c = 1.19$ for \cite{fall2015} and \cite{hermes2016}. $\Delta=0$ for all experiments.} \label{tab:experiment}
		\begin{tabular}{  c  c  c  c  c  c  c  } 
			Fit & $\phi_c$ & $\phi_j$ & $a_0$ & $a_\infty$ & $\tau^*$ [Pa] & $\eta_0$ [mPa$\cdot$s]\\
			\midrule
			\cite{dheane1993} & 0.57 & 0.65 & 1.08 & 1.08 & 5.21 & 56\\
			\cite{frith1996} & 0.548 & 0.642 & 1.08 & 1.08 & 17.93 & 130\\
			\cite{guy2015}(a) & 0.56 & 0.617 & 1.577 & 1.577 & 423.0 & 2.4\\
			\cite{guy2015}(b) & 0.563 & 0.610 & 1.09 & 1.538 & 3.5 & 2.4\\
			\cite{fall2015} & 0.38 & 0.454 & 1.00 & 1.00 & 0.116 & 0.89\\
			\cite{hermes2016} & 0.436 & 0.535 & 0.8 & 0.8 & 5.35 & 6\\
			\bottomrule
		\end{tabular}
	\end{center}
\end{table}

\begin{table}[h]
	\small
	\begin{center}
		\caption{\small Parameters of $\hat{\eta}_B(\phi)$ for curves in figure \ref{fig:rheochaoticflows}.} \label{tab:experiment_etaB}
		\begin{tabular}{  c  c  c  c  c  } 
			Fit & $\varphi_1$ [(Pa$\cdot$s)$^{-1}$] & $\alpha_1$ & $\varphi_2$ [(Pa$\cdot$s)$^{-1}$] & $\alpha_2$ \\
			\midrule
			\cite{dheane1993} & $8.83 \times 10^6$ & 6 & 0 & 0\\
			\cite{frith1996} & $1.79 \times 10^{11}$ & 12 & $4.5$ & 0.1 \\
			\cite{guy2015}(a) & $3 \times 10^6$ & 5 & $9 \times 10^{-2}$ & 0.1\\
			\cite{guy2015}(b) & $1.0 \times 10^{6}$ & 6 & 0 & 0\\
			\cite{fall2015} & $4.0 \times 10^4$ & 3 & 0 & 0\\
			\cite{hermes2016} & $1.86 \times 10^{4}$ & 4 & $9.3 \times 10^{-3}$ & 0.1\\
			\bottomrule
		\end{tabular}
	\end{center}
\end{table}


\subsection*{Complete, Time-Dependent Expression of Model}
The steady state measurement of DST and CST is only one part of the puzzle of repulsive grain suspensions. As noted in \cite{waitukaitis2012}, a fully coupled constitutive model is necessary to capture the observed dynamic behavior of these mixtures as epitomized by the `running on oobleck' effect. This effect describes the ability of a person to run over the surface of a mixture of standard cornstarch and water without sinking (within a certain range of volume fractions); however, if a person walks over the same mixture, they rapidly sink to the bottom.

In the next section, we will show that the model proposed in this work is capable of capturing this behavior; however, we must first express the time-dependent, three-dimensional form of our model. The complete set of governing equations (many of which are unchanged from the model developed in \cite{baumgarten2018}) can be found in the Supplemental Information. Here, we focus only on the granular phase stress components $\tilde{\sigma}_{ij}$ which obey a Maxwell-like model. We assume that the deformation rate tensor, $D_{ij} = \frac{1}{2}(\partial v_{si}/ \partial x_j + \partial v_{sj}/ \partial x_i)$, can be decomposed into an elastic part $\tilde{D}^e_{ij}$ and plastic part $\tilde{D}^p_{ij}$ with ${v_s}_i$ the components of the granular phase velocity. This allows us to express the rate of change of the granular phase stress $\tilde{\sigma}_{ij}$ as a function of the elastic deformation rate, the granular shear modulus $G$ and bulk modulus $K$.
%
%
%
%
%
%
%
%
The components of the plastic part of the deformation rate tensor can be expressed using the form given in \cite{baumgarten2018} as follows,
\begin{equation} \label{eqn:plastic}
\tilde{D}^p_{ij} = \frac{\dot{\bar{\gamma}}^p}{2\bar{\tau}} (\tilde{\sigma}_{ij} + \tilde{p}\delta_{ij}) + \tfrac{1}{3}(\beta \dot{\bar{\gamma}}^p + \dot{\xi}_1 + \dot{\xi}_2)\delta_{ij},
\end{equation}
with,
\begin{equation}
\bar{\tau} = \sqrt{\tfrac{1}{2}(\tilde{\sigma}_{kl} + \tilde{p}\delta_{kl})(\tilde{\sigma}_{kl} + \tilde{p}\delta_{kl})}, \qquad
\tilde{p} = -\tfrac{1}{3} \tilde{\sigma}_{kk},
\end{equation}
$\delta_{ij}$ the Kronecker delta function, and $\dot{\bar{\gamma}}^p$ the scalar (equivalent) plastic shear rate that generalizes $\dot{\gamma}$ to unsteady, three-dimensional flows.
The expression in \eqref{eqn:plastic} captures the two mechanisms of granular flow: shearing and dilation. The second term in \eqref{eqn:plastic} (representing dilation of the granular phase) captures the effects of Reynolds dilation ($\beta \dot{\bar{\gamma}}^p$), pure expansion ($\dot{\xi}_1$), and pure compaction ($-\dot{\xi}_2$). $\beta$, $\dot{\bar{\gamma}}^p$, and $\dot{\xi}_2$, are functions of the granular stress $\tilde{\sigma}_{ij}$, $a$, and $\phi_m$; $\dot{\xi}_1$ is a function of inter-granular cohesion only. Implementation of the changes made in our proposed model is achieved by replacing $a$ and $\phi_m$ in the functions for $\beta$, $\dot{\bar{\gamma}}^p$, and $\dot{\xi}_2$ with \eqref{eqn:phim_func} and \eqref{eqn:a_func}, and substituting $\dot{\bar{\gamma}}^p$ in place of $\dot{\gamma}$ in \eqref{eqn:c_model}. 

\section{Results} \label{sec:results}
In this section, we show that the fully three-dimensional, time-dependent, two-phase form of our model accurately reproduces the observed transient behavior of cornstarch-water mixtures in annular shear and impact experiments as well as the `running on oobleck' effect. To perform this analysis, we implement our model in the numerical framework shown in \cite{baumgarten2018}. This numerical framework is an adaptation of the material point method (MPM) for simulating mixtures as two overlapping continua. This version of MPM is very similar to the method shown in \cite{bandara2015} and differs from the original method described in \cite{sulsky1994} in that each continuum phase of the mixture is represented by independent sets of material point tracers. These material point tracers act as quadrature points for solving the weak form of the momentum balance equations on a static background simulation grid. In addition, these tracers move with the continua that they represent. To understand the results shown in this work, it suffices to understand that the velocities and strain-rates of the continua are represented on the nodes of the simulation grid while displacements and stresses are represented on the material point tracers.

{

As in \cite{baumgarten2018}, we assume that the granular phase of the mixture is \textit{elastically stiff} (i.e.\ elastic moduli $G$ and $K \gg \bar{\tau}$) so that any significant deformation of the material comes from plastic flow ($\dot{\bar{\gamma}}^p$). In all of the simulations reported in this work, we choose values of the fluid bulk modulus and the moduli $G$ and $K$ to be large enough such that any effects of elasticity do not alter the flow solution (to do this, we require that all elastic deformations are smaller than 1\%) but not so large that they significantly impact the quality of the numerical calculations. In this way, we ensure that our results are representative of elastically stiff mixtures while also avoiding well known numerical artifacts in MPM (see kinematic locking in \cite{mast2012} and the ringing instability in \cite{gritton2014}) that affect the smoothness of the spatial stress field. Further discussion of this point can be found in the Supplemental Information along with a brief note about the validity of the elastically stiff assumption.
}

\subsection*{Dynamic Shear Jamming}
\begin{figure}[!h]
	\centering
	\includegraphics[scale=0.43]{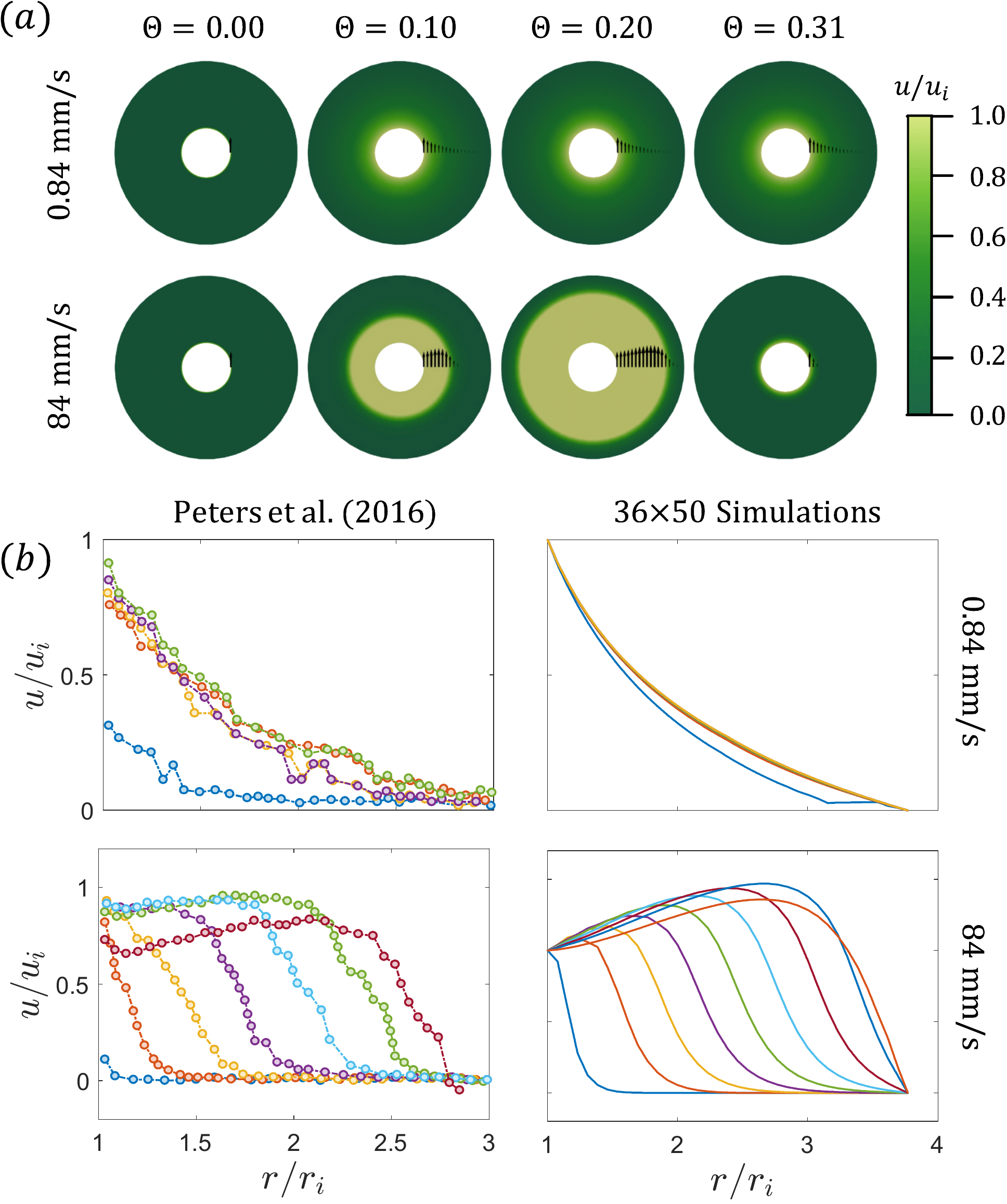}
	\caption{\footnotesize (a) Comparison of surface velocity field for annular shear simulations (see table \ref{tab:peters}) after 0, 0.1, 0.2, and 0.31 radians of bob (inner wall) rotation for $u_i = 0.84$ mm/s and $u_i = 84$ mm/s. The $u_i = 0.84$ mm/s simulation rapidly approaches the diffusive flow profile of Newtonian fluids in annular shear (as observed in \cite{peters2016}). However, the $u_i = 84$ mm/s simulation shows rapid growth of a rigidly rotating block of mixture; when the edge of this block reaches the outer edge of the annular shear cell, all flow ceases. (b) Comparison of experimental surface velocity profiles taken at even intervals of rotation from \cite{peters2016} to velocity profiles in simulations on 36 $\times$ 50 axisymmetric grid. Note, simulations do not account for wall slip.} \label{fig:peters}
\end{figure}

Here we show that our model can capture the reported propagation of `shear jamming fronts' in annular shear experiments; in particular, we are interested in the starkly different mixture responses reported at different applied shearing rates. One example of this phenomena can be found in \cite{peters2016} which reports several experiments involving density matched mixtures of cornstarch, water, gylcerol, and CsCl. In the annular shear cell described in that work, an applied inner wall velocity of $u_i=0.84$ mm/s produces shearing flow that rapidly approaches the diffusive flow profile expected from a Newtonian fluid; however, when the applied inner wall velocity is multiplied by a factor of 100, the mixture begins `jamming' at the inner wall. This `jammed' region is characterized by nearly rigid rotation (see figure \ref{fig:peters}) and a well defined boundary that propagates outward with a near constant rate.

We recreate the reported experiment with our continuum model using an axisymmetric form of MPM on a 36$\times$50 element Cartesian grid  with 1mm$\times$1mm resolution. Our model is calibrated to the experiments using the steady flow parameters for the cornstarch-water/glycerol experiments of \cite{hermes2016} (see tables \ref{tab:experiment} and \ref{tab:experiment_etaB}) and material parameters as in table \ref{tab:peters}. The only parameters that were tuned to fit experimental observations were $K_0$ and $\phi_0$. The values chosen for $\eta_0$, $d$, and $\rho$ are informed by the materials used in the mixtures ($d$ in all simulations is chosen to be 5.85 $\mu$m; see \cite{stasiak2013}). $K_3$ and $K_4$ are chosen according to the findings of \cite{baumgarten2018} (4.715 and 0, respectively). The simulation results in figure \ref{fig:peters} show excellent agreement with experimental observations neglecting wall slip effects that were not implemented in our model.
\begin{table}[h]
	\small
	\caption{\small Material parameters for simulations shown in figure \ref{fig:peters}.} \label{tab:peters}
	\begin{center}
		\begin{tabular}{  c  c  c  c  c   c  } 
			$\phi_0$ & $\eta_0$ [mPa$\cdot$s] & $\rho$ [kg/m$^3$] & $G$ [Pa] & $K$ [Pa] & $K_0$\\
			\midrule
			0.505 & 10 & 1620 & $3.8\times10^5$ & $8.3\times10^5$ & 0.06\\
			\bottomrule
		\end{tabular}
	\end{center}
\end{table}

\subsection*{Impact Activated Solidification}
In this section, we show that our model accurately captures the solidification of constarch-water/glycerol mixtures under impact. Similar to the `shear jamming' observed in the previous section, this `impact activated solidification' results in two strikingly different material responses. At high impact speeds the mixture responds like a solid, sometimes causing the impactor to bounce (see \cite{waitukaitis2012}); at low impact speeds the mixture responds like a fluid, flowing around the intruder with ease. In addition, as predicted in \cite {waitukaitis2012} and directly observed in \cite{peters2014} and \cite{han2016}, at high impact speeds, a `solid plug' forms at the impact site and propagates outward (another type of `jamming front').

\begin{figure}[!h]
	\centering
	\includegraphics[scale=0.35]{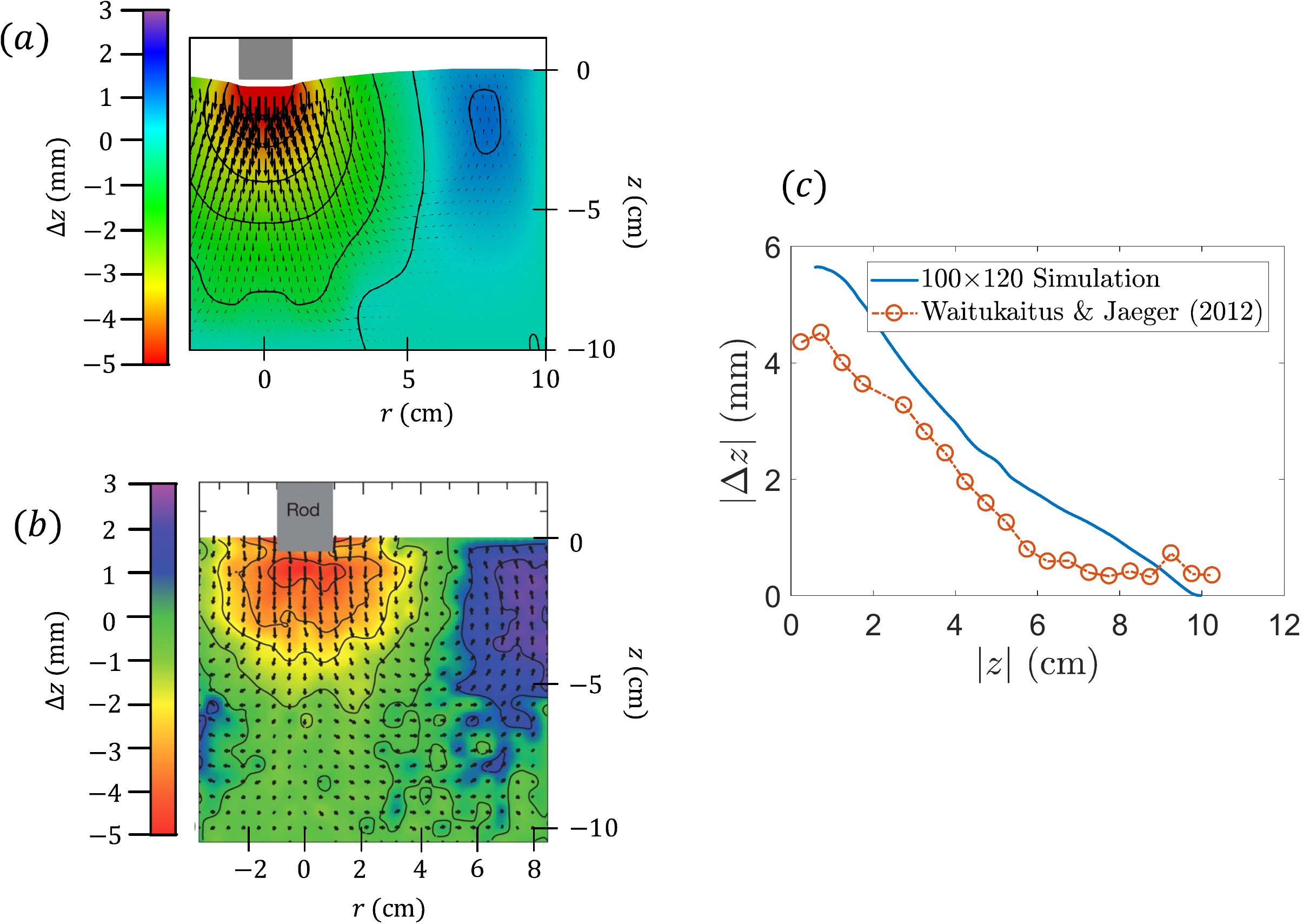}
	\caption{\footnotesize (a) Plot of $z$-displacement field and contours for simulated cornstarch-water/glycerol mixture ($\eta_0 = 7$ cP) after $\Delta t = 60$ ms for aluminum rod impacting 10 cm deep, radially symmetric mixture ($r_{\text{mixture}}=10$ cm; $r_{\text{rod}}=0.95$ cm) at 1.0 m/s. Simulation is performed on 100$\times$120 element grid with 1mm$\times$1mm resolution. The red and yellow region near the impactor represents a region of `jammed' material co-moving with the rod. (b) Plot of experimental $z$-displacement field taken from \cite{waitukaitis2012} (image from \cite{waitukaitis2012}). (c) Comparison of displacement $|\Delta z|$ vs. depth $z$ directly below the impacting rod after $\Delta t=60$ ms for 100$\times$120 element simulation and results reported in \cite{waitukaitis2012}. The roughly linear profile observed in the simulated results and in the experiment is indicative of the steadily growing, `solid plug' of material below the rod.} \label{fig:waitukaitus3}
\end{figure}

\begin{figure}[!h]
	\centering
	\includegraphics[scale=0.45]{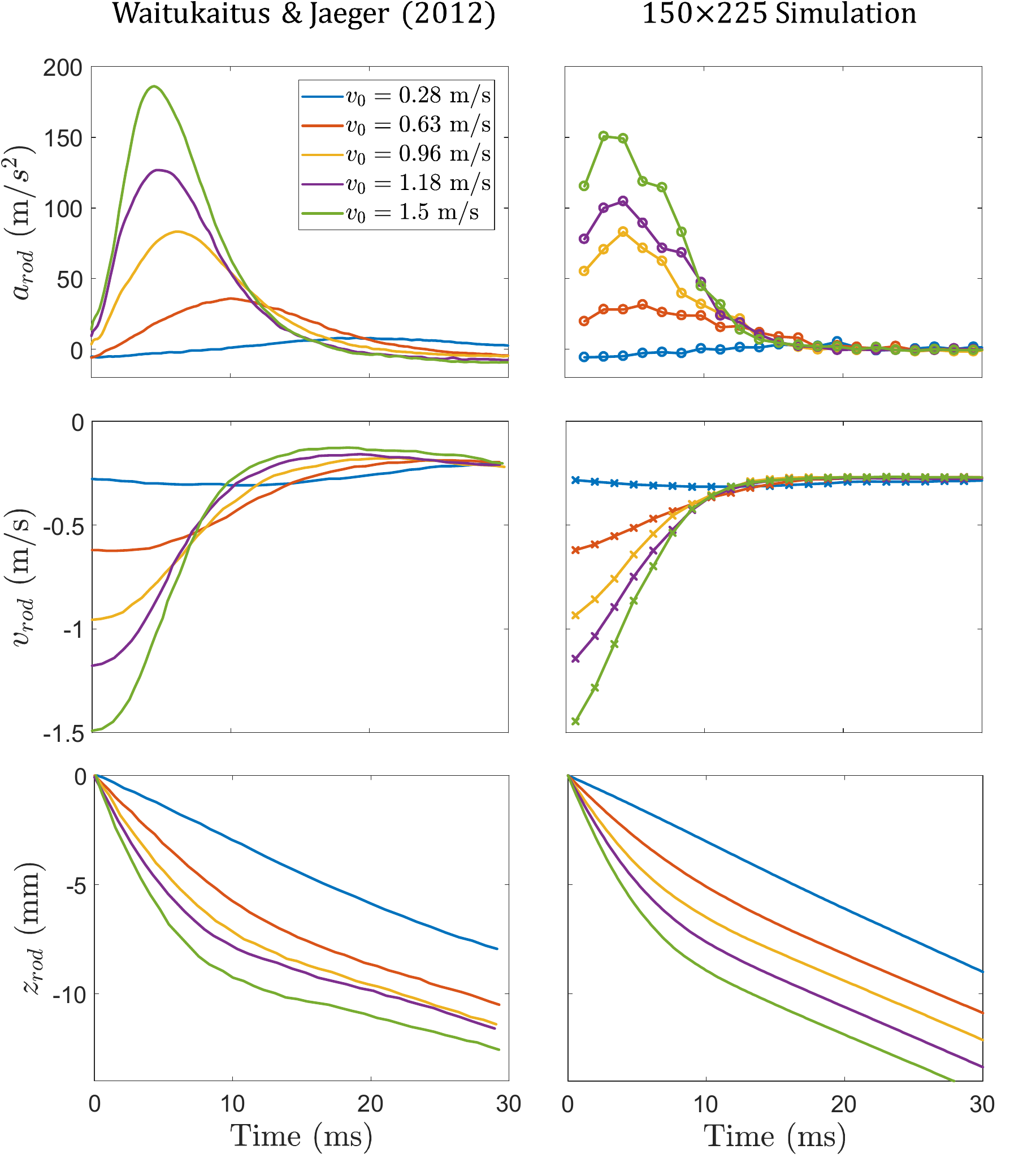}
	\caption{\footnotesize Comparison of rod motion (acceleration $a_{\text{rod}}$, velocity $v_{\text{rod}}$, and position $z_{\text{rod}}$) from large container experiments ($\eta_0 = 1$ cP) in \cite{waitukaitis2012} to simulated rod dynamics from axisymmetric, 150$\times$225 element Cartesian grid simulation. Several initial rod velocities $v_0$ are shown (see legend). {Simulated rod velocity and acceleration profiles are obtained from the simulated position profiles using a first-order finite difference scheme. The length of the finite difference stencil is chosen to remove the effects of elastic waves rebounding off of the simulation boundaries.}} \label{fig:waitukaitus4}
\end{figure}

We begin analyzing the response of our (fully dynamic) model under impact by recreating several of the experiments reported in \cite{waitukaitis2012} where an aluminum rod is launched at a mixture of cornstarch, water, and glycerol. Measurement of the rod motion was taken visually and with an accelerometer, and mixture displacement fields were visualized using X-ray imaging. We calibrate our model to the experiments using the steady flow parameters for \cite{hermes2016} in tables \ref{tab:experiment} and \ref{tab:experiment_etaB} and material parameters in table \ref{tab:waitukaitus} ($d$, $K_3$, and $K_4$ are chosen as in the previous section). The simulations are performed using axisymmetric MPM on Cartesian grids with 1mm$\times$1mm resolution. We only tune $K_0$ to accurately reproduce the displacement fields (see figure \ref{fig:waitukaitus3}) and rod dynamics (see figure \ref{fig:waitukaitus4}) reported in \cite{waitukaitis2012}. {{\color{blue}}By utilizing the full continuum momentum balance, our model naturally captures observed added mass effects from material inertia.}

\begin{table}[h]
	\small
	\caption{\small Material parameters for simulations shown in figures \ref{fig:waitukaitus3} and \ref{fig:waitukaitus4}.} \label{tab:waitukaitus}
	\begin{center}
		\begin{tabular}{  c  c  c  c  c  c  } 
			$\eta_0$ [mPa$\cdot$s] & $\rho_s$ [kg/m$^3$] & $\rho_f$ [kg/m$^3$] & $G$ [Pa] & $K$ [Pa] & $K_0$\\
			\midrule
			7 or 1 & 1620 & 1000 & $3.8\times10^7$ & $8.3\times10^7$ & 1.0\\
			\bottomrule
		\end{tabular}
	\end{center}
\end{table}

The behavior of our model during impact shows remarkable similarity to experiments. The stiff response of the mixture at high impact speeds (as shown by the acceleration curves in figure \ref{fig:waitukaitus4}) is contrasted by the steady sinking that is observed once the rods have slowed (as shown by the velocity curves in figure \ref{fig:waitukaitus4}). The X-ray imaging performed in \cite{waitukaitis2012} allowed observation of the final displacement field within the mixture (see figure \ref{fig:waitukaitus3}) and the results are indicative of a `solidification front' propagating during impact.

We continue the analysis of our model during impact by recreating the controlled intrusion experiments described in \cite{han2016} with a focus on measuring the rate of propagation of the `solidification front' as the intruder enters the mixture. In these experiments, a rod was driven with a constant velocity $v_0$ into a density matched cornstarch-water/glycerol/CsCl mixture, and the internal flow of the mixture was visualized using high speed ultrasound imaging. We calibrate our model to the experiments using the parameters for \cite{hermes2016} found in tables \ref{tab:experiment} and \ref{tab:experiment_etaB} and material parameters in table \ref{tab:han} ($d$, $K_3$, and $K_4$ are chosen as before). The simulations are run using axisymmetric MPM on a 50$\times$35 element Cartesian grid with 1mm$\times$1mm resolution.

\begin{table}[h]
	\small
	\caption{\small Material parameters for simulations shown in figures \ref{fig:han2} and \ref{fig:han4}.} \label{tab:han}
	\begin{center}
		\begin{tabular}{  c  c  c  c  c  c  } 
			$\eta_0$ [mPa$\cdot$s] & $\rho_s$ [kg/m$^3$] & $\rho_f$ [kg/m$^3$] & $G$ [Pa] & $K$ [Pa] & $K_0$\\
			\midrule
			4.6 & 1620 & 1620 & $3.8\times10^7$ & $8.3\times10^7$ & 0.2\\
			\bottomrule
		\end{tabular}
	\end{center}
\end{table}

\begin{figure}[!h]
	\centering
	\includegraphics[scale=0.52]{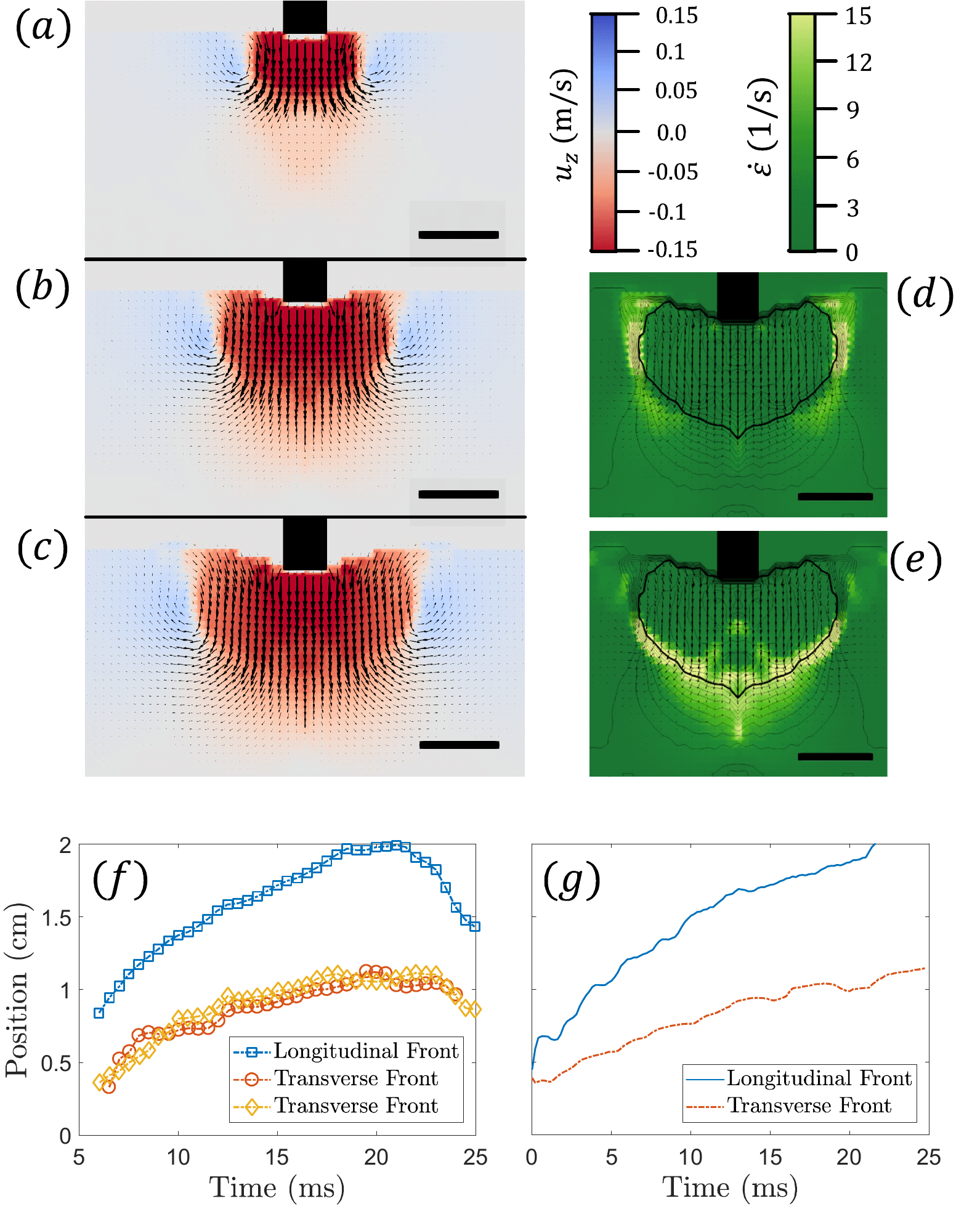}
	\caption{\footnotesize Visualization of flow field and front motion from axisymmetric MPM simulation of intruder. Black bars provide 1cm length scale. The driving velocity $v_0$ of the intruder is 0.175 m/s and mixture volume fraction $\phi$ is 0.47 for results shown in (a) through (e). The driving velocity $v_0$ is 0.2 m/s and mixture volume fraction $\phi$ is 0.46 for results shown in (f) and (g). (a) Flow field colored by vertical component of velocity $u_z$ at $t=6.2$ ms. (b) Flow field colored by $u_z$ at $t=13.2$ ms. (c) Flow field colored by $u_z$ at $t=20$ ms. (d) Visualization of strain rate $\dot{\varepsilon} = |D_{rz}|$ at $t = 20$ ms with contours of velocity $u_z$ ($0.5 v_0$ in bold). (e) Visualization of strain rate $\dot{\varepsilon} = |D_{zz}|$ at $t = 20$ ms. (f) Experimental measurement of front position (defined by $u_z = 0.5 v_0$) in the longitudinal and transverse directions from \cite{han2016}. (g) Simulated front position in the longitudinal and transverse directions.}
	\label{fig:han2}
\end{figure}

\begin{figure}[!h]
	\centering
	\includegraphics[scale=0.75]{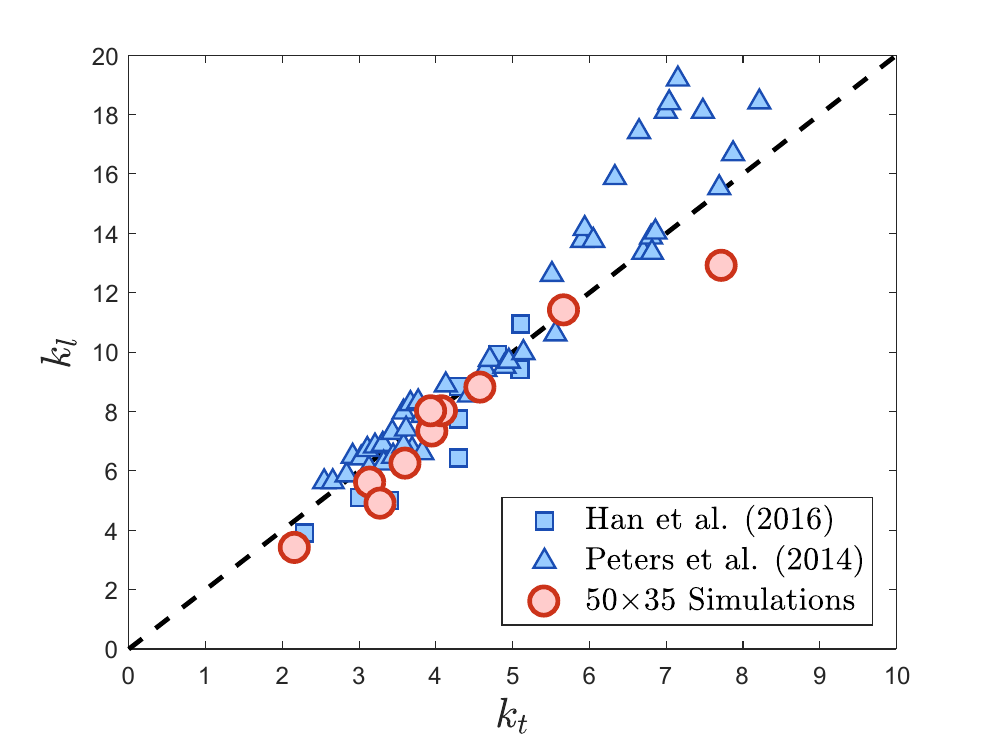}
	\caption{\footnotesize Scatter plot of front propagation factors $k_l$ and $k_t$ from \cite{han2016} (blue squares), \cite{peters2014} (blue triangles), and from several axisymmetric MPM simulations on a 50$\times$35 element Cartesian grid (red circles). The black dashed line is the proposed front propagation ratio reported in \cite{han2016}, $k_l/k_t \approx 2$. The simulation results shown here were run at volume fractions $\phi$ of [0.46, 0.47, 0.48, 0.49, 0.50] and intruder velocities $v_0$ of [200 mm/s, 500 mm/s]. Measurement of simulated front propagation speed is performed using the method described in \cite{han2016}.}
	\label{fig:han4}
\end{figure}

Tuning $K_0$, we are able to closely match the reported flow fields (see figure \ref{fig:han2}) from \cite{han2016}. As the intruder enters the mixture, a `solid plug' of material forms below the intruder and grows outward in all directions. This `solid plug' is characterized by nearly rigid vertical motion with a well defined boundary that grows with near constant rate. The rate of propagation of this boundary in the transverse $v_{ft}$ and longitudinal $v_{fl}$ directions is often quite different (as shown in figure \ref{fig:han2}(f) and \ref{fig:han2}(g)); however, these rates appear to obey a particular scaling rule that is independent of volume fraction: $(v_{fl}-v_0)/v_{ft} \approx 2$. Remarkably, our simple three-dimensional model naturally captures this reported scaling (see figure \ref{fig:han4}; results are shown in terms of the transverse and longitudinal front propagation factors $k_t = v_{ft}/v_0$ and $k_l = v_{fl}/v_0 - 1$, respectively). This close agreement of model predictions to experimental observations is unmatched by any other models in the literature.

\begin{figure*}[!ht]
	\centering
	\includegraphics[scale=0.42]{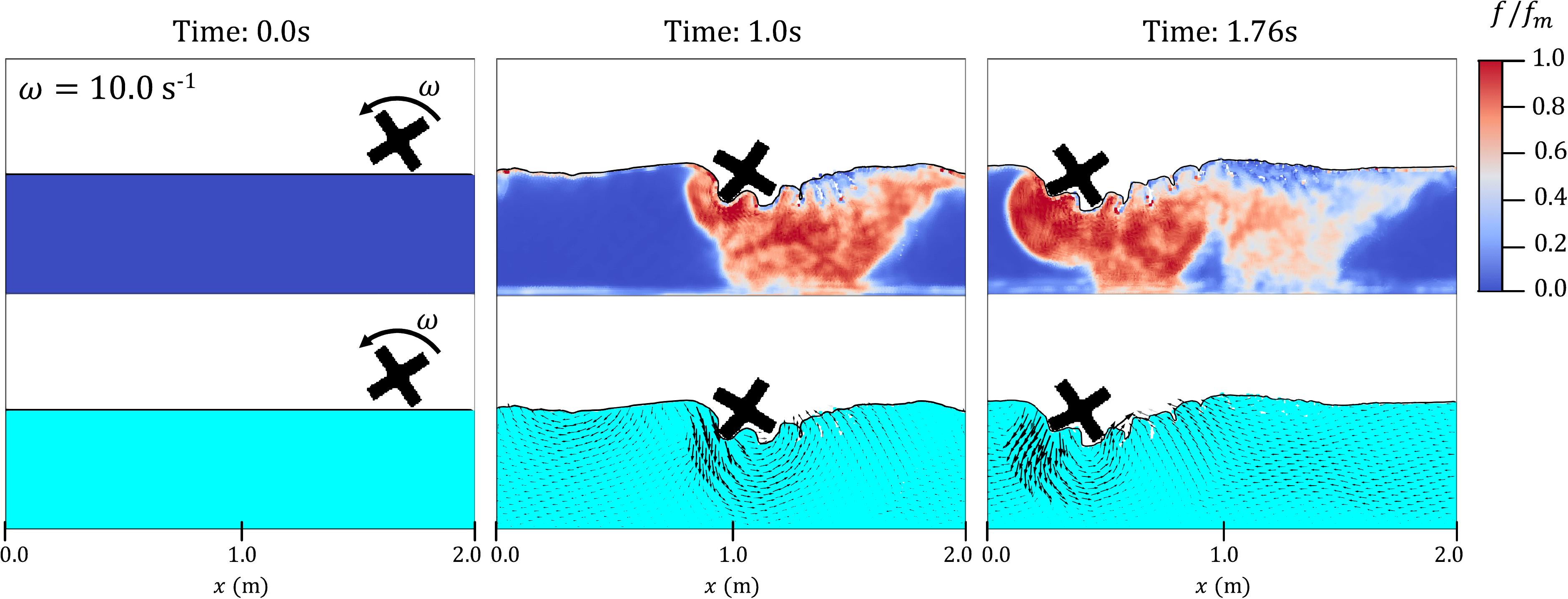}
	\caption{\footnotesize Visualization of structural field $f$ field and velocity vector field in simulated cornstarch-fluid mixture for wheel driven at $\omega = 10.0$ s$^{-1}$. The value of $f$ is used to color the material point tracers representing the cornstarch phase of the mixture. The black vectors representing the velocity field are drawn with component lengths $l_i = 1.215 v_i \sqrt{R/g}$. A black contour line is added to aid in visualizing the free surface of the mixture and is associated with the volume fraction $\phi  = 0.1$ as represented on the background grid. This contour does not exactly represent the free surface, as accumulated numerical errors cause disagreement between material point volume fractions and effective background grid volume fractions; these numerical errors are caused by topological changes along the surface of the material but do not significantly affect the qualitative behavior reported.}
	\label{fig:wheel_fast}
\end{figure*}

\begin{figure*}[!ht]
	\centering
	\includegraphics[scale=0.42]{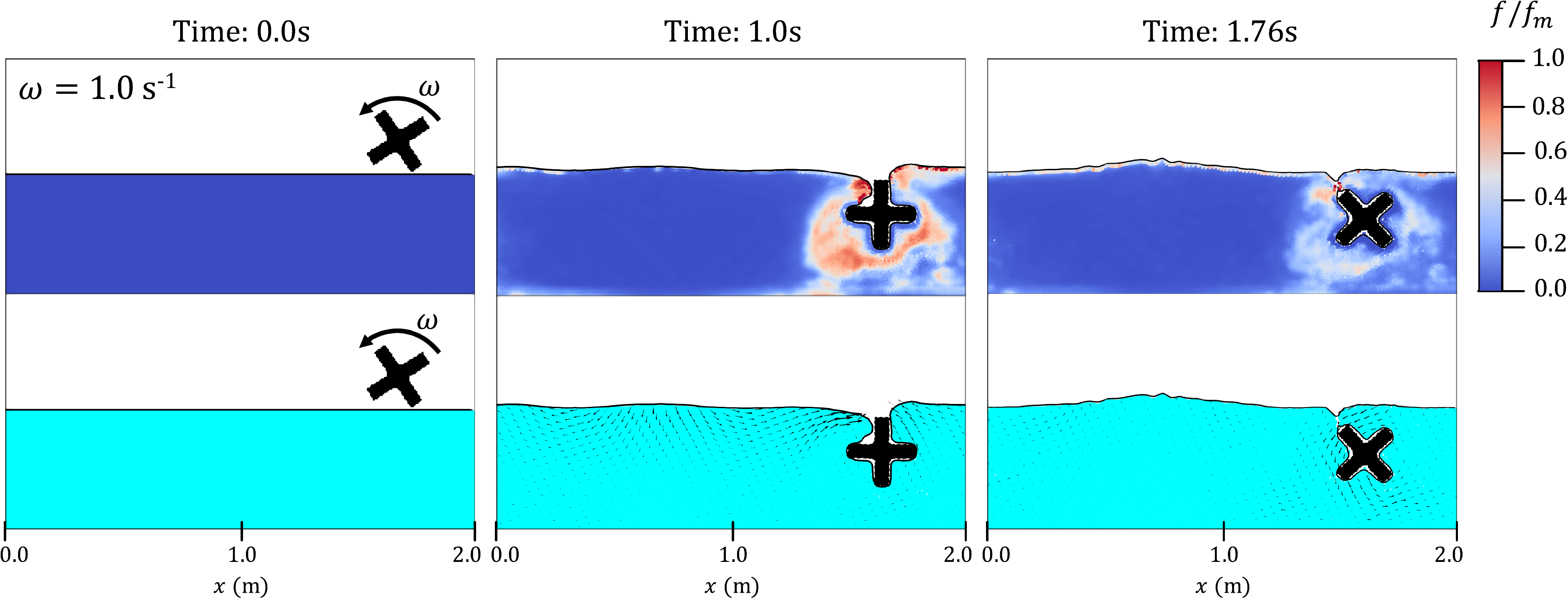}
	\caption{\footnotesize Visualization of structural field $f$ field and velocity vector field in simulated cornstarch-fluid mixture for wheel driven at $\omega = 1.0$ s$^{-1}$. The value of $f$ is used to color the material point tracers representing the cornstarch phase of the mixture. The black vectors representing the velocity field are drawn with component lengths $l_i = 1.215 v_i \sqrt{R/g}$. A black contour line is added to aid in visualizing the free surface of the mixture and is associated with the volume fraction $\phi  = 0.1$ as represented on the background grid.}
	\label{fig:wheel_slow}
\end{figure*}

\subsection*{Running on Oobleck}
In this section we demonstrate that our model can capture the `running on oobleck' effect by simulating a spoked elastic wheel driving over the surface of a density matched cornstarch-fluid mixture similar to those described above. The simulation is run using two-dimensional MPM on a 150$\times$75 element (2m$\times$1m) periodic Cartesian grid using the steady flow parameters for \cite{hermes2016} in tables \ref{tab:experiment} and \ref{tab:experiment_etaB} with material parameters in table \ref{tab:wheel}. ($d$, $K_3$, and $K_4$ are chosen as in the previous sections). The four spoked elastic wheel used in this simulation has radius 15cm and density 3000 kg/m$^3$ and is driven with constant rotation rate $\omega$ starting on the mixture's surface (see figures \ref{fig:wheel_fast} and \ref{fig:wheel_slow}). 

\begin{table}[h]
	\small
	\caption{\small Material parameters for simulations shown in figures \ref{fig:wheel_fast} and \ref{fig:wheel_slow}.} \label{tab:wheel}
	\begin{center}
		\begin{tabular}{  c  c  c  c  c  c  } 
			$\eta_0$ [mPa$\cdot$s] & $\rho_s$ [kg/m$^3$] & $\rho_f$ [kg/m$^3$] & $G$ [Pa] & $K$ [Pa] & $K_0$\\
			\midrule
			1.0 & 1620 & 1620 & $3.8\times10^5$ & $8.3\times10^5$ & 1.0\\
			\bottomrule
		\end{tabular}
	\end{center}
\end{table}

As shown in figures \ref{fig:wheel_slow} and \ref{fig:wheel_fast} respectively, the wheel driving at $\omega = 1.0$ s$^{-1}$ will rapidly sink while the wheel driving at $\omega = 10.0$ s$^{-1}$ will run along the surface. This difference in behavior is caused by the same mechanism explored in the previous section. In the slow case ($\omega = 1.0$ s$^{-1}$), the spokes of the wheel are traveling too slowly to produce a solid-like response; the entire wheel slips into the mixture as it would in a fluid. However, in the fast case ($\omega = 10.0$ s$^{-1}$), when the spokes of the wheel impact the mixture surface, they cause rapid solidification of the surrounding mixture --- see the spacial $f$ field in figures \ref{fig:wheel_fast} and \ref{fig:wheel_slow} (plotted using a post-processing routine from \cite{dunatunga2015}). This allows that spoke to support the weight of the wheel until the next spoke strikes the surface further on; the wheel runs along the top of the mixture.

\section{Conclusion} \label{sec:conclusion}
The model proposed in this work shows remarkable accuracy in predicting the behavior of repulsive particle suspensions in both steady and unsteady flows. By combining a model for the evolution of granular micro-structure with the governing equations for fluid-sediment mixtures presented in \cite{baumgarten2018}, we have created a single constitutive model for chemically stable, hard, frictional particles suspended in viscous fluids. In addition, implementation of this model in the numerical framework proposed in \cite{baumgarten2018} allows simulation of the coupled behavior of particles and fluid through a wide range of particle sizes, volume fractions, and flow regimes (including independent motion of each phase and interaction with solid bodies). We have demonstrated that this model captures the landmark features of these mixtures including (i) DST and CST, (ii) propagation of `shear jamming fronts', (iii) the propagation of `impact activated jamming fronts', and (iv) the `running on oobleck' effect.

There are a number of areas remaining for further research and improvement of this model. In this work, we have focused on the time-average behavior of rheologically chaotic flows; however, the basic model may be used to predict the time-accurate behavior of these flows through a fabric tensor contribution to the evolution of $f$ (as noted in \cite{dheane1993} and \cite{stickel2005}). In addition, future work will need to address the dependence of $\tau^*$, $\eta_B$, and $K_0$ on changes in particle and fluid material properties such as $\eta_0$, $d$, $\rho$, and $\theta$ (with $\theta$ temperature; see \cite{warren2015}) and the additional dependence of $K_0$ on stress scale $\bar{\tau}$ and shearing rate $\dot{\gamma}$. {
In the Supplemental Information, we describe a simple relaxation experiment which could help determine these dependencies. Numerically, we have shown that our model can be implemented in MPM (see \cite{baumgarten2018}) and accurately reproduce the time-dependent behavior of these highly non-linear mixtures in non-trivial geometries. Although our initial tests are promising, further work should also be done to curb known issues in the numerical framework such as kinematic locking (see \cite{mast2012}) and the ringing instability (see \cite{gritton2014}), which can cause spurious artifacts in certain parameter ranges if not dealt with carefully.
}

\bibliography{references}

\onecolumn
\section{Supplemental Information}

\subsection*{Complete Table of Equations}
The complete, three-dimensional, two-phase model proposed in this work is an extension of the model presented in \cite{baumgarten2018}. As described in the main document, this model changes the definition of the dilation parameters $a$ and $\phi_m$ from material constants (as in \cite{baumgarten2018}) to functions of the granular micro-structure $f$. A summary of the equations that define the complete model can be found in table \ref{tab:governing_equations}.

These equations govern the behavior of the two independent continua that form the mixture: the granular phase and the fluid phase. The mass conservation rules govern the evolution of the fluid phase \textit{true} density $\rho_f$ and \textit{effective} density $\bar{\rho}_f = n \rho_f$ as well as the granular phase \textit{effective} density $\bar{\rho}_s = \phi \rho_s$. Here we assume that $\rho_s$ is constant (incompressible grains) and define $n$, the mixture porosity, as equal to $(1-\phi)$. The two momentum balance equations describe the evolution of the components of the fluid and granular phase velocities ${v_f}_i$ and ${v_s}_i$ according to gradients in the fluid phase pore pore pressure $p_f$, components of the effective granular stress tensor $\tilde{\sigma}_{ij}$, and components of the fluid phase shear stress tensor ${\tau_f}_{ij}$. In addition, these rules also define the components of the inter-phase Darcy's Law drag that acts between the two material phases, ${f_d}_i$; the scalar function $\hat{F}(\phi, \textrm{Re})$ is defined in \cite{baumgarten2018} and taken from \cite{beetstra2007}.

The fluid phase stresses are modeled using a viscous, barotropic constitutive model which allows for weak compressibility and accounts for Einstein's effective viscosity model for dilute suspensions.

The granular phase effective stress is modeled as an elastic-plastic solid. Through careful definition of the components of the plastic flow rate tensor $\tilde{D}^p_{ij}$ (given by the scalar rates $\dot{\bar{\gamma}}^p$, $\dot{\xi}_1$ and $\dot{\xi}_2$), this model can accurately describe stresses in the granular skeleton of a mixture at rest, in steady flow, and all regimes in between. In addition, the form of the plastic flow rate tensor in table \ref{tab:governing_equations} captures the different modes of deformation which a granular material can undergo: pure shear ($\dot{\bar{\gamma}}^p$), shear dilation ($\beta \dot{\bar{\gamma}}^p$), pure dilation ($\dot{\xi}_1$), and pure compaction ($-\dot{\xi}_2$). The scalar flow rates $\dot{\bar{\gamma}}^p$, $\dot{\xi}_1$ and $\dot{\xi}_2$ are themselves functions of the granular phase effective stress and are defined implicitly by the yield functions $f_1$, $f_2$, and $f_3$ respectively. These yield functions are written such that the granular phase captures the $\mu(I)$ dry inertial rheology from \cite{jop2006}, the $\mu(I_v)$ viscous inertial rheology from \cite{boyer2011}, and the $\mu(I_m, I_v)$ mixed inertial rheology from \cite{amarsid2017}.

\begin{table}[h]
	\small
	\begin{center}
		\caption{\small Summary of equations for full three-dimensional model.} \label{tab:governing_equations}
		\def\arraystretch{2}
		\begin{tabular}{ l  c } 
			\textbf{Rule}&  \textbf{Expression}\\ 
			\midrule
			Granular Phase Mass Conservation & $\displaystyle\frac{D^s \bar{\rho}_s}{Dt} + \bar{\rho}_s \sum_{i=1}^3 \frac{\partial {v_s}_i}{\partial x_i} = 0$\\
			Fluid Phase Mass Conservation &  $\displaystyle\frac{D^f \bar{\rho}_f}{Dt} + \bar{\rho}_f \sum_{i=1}^3 \frac{\partial {v_f}_i}{\partial x_i} = 0$\\
			Fluid Phase True Density &  $\displaystyle\frac{n}{\rho_f}\frac{D^f \rho_f}{Dt} = -\sum_{i=1}^3 \frac{ \partial ((1-n){v_s}_i + n {v_f}_i)}{\partial x_i}$\\
			Granular Phase Momentum Balance &  $\displaystyle\bar{\rho}_s\frac{D^s {v_s}_i}{Dt} = \bar{\rho}_s {g}_i - {f_d}_i - (1-n)\frac{\partial p_f}{\partial x_i} + \sum_{i=1}^3 \frac{\partial \tilde{\sigma}_{ij}}{\partial x_j}$\\
			Fluid Phase Momentum Balance & $\displaystyle\bar{\rho}_f\frac{D^f {v_f}_i}{Dt} = \bar{\rho}_f {g}_i + {f_d}_i - n \frac{\partial p_f}{\partial x_i} + \sum_{i=1}^3 \frac{\partial {\tau_f}_{ij}}{\partial x_j}$\\
			Darcy's Drag Law & $\displaystyle {f_d}_i = \frac{18\phi(1-\phi)\eta_0}{d^2} \ \hat{F}(\phi,\mathrm{Re}) \ ({v_s}_i-{v_f}_i)$\\
			Fluid Phase Pore Pressure & $\displaystyle p_f = \kappa \ln\bigg(\frac{\rho_f}{\rho_{0f}}\bigg)$\\
			Fluid Phase Shear Stress & $\displaystyle {\tau_f}_{ij} = \eta_0 \big(1 + \tfrac{5}{2}\phi\big) \bigg( \frac{\partial {v_f}_i}{\partial x_j} + \frac{\partial {v_f}_j}{\partial x_i} - \frac{2}{3} \sum_{k=1}^{3} \frac{\partial {v_f}_k}{\partial x_k} \delta_{ij}  \bigg) $\\
			Granular Phase Effective Stress & $\displaystyle
			\dot{\tilde{\sigma}}_{ij} = 2 G (D_{ij} - \tilde{D}^p_{ij}) + (K - \tfrac{2}{3}G)(D_{kk} - \tilde{D}^p_{kk})\delta_{ij} + W_{ik}\tilde{\sigma}_{kj} - \tilde{\sigma}_{ik}W_{kj}
			$\\
			Granular Phase Strain-Rate and Spin Tensors &
			$\displaystyle D_{ij} = \frac{1}{2} \bigg(\frac{\partial {v_s}_i}{\partial x_j} + \frac{\partial {v_s}_j}{\partial x_i} \bigg), \quad W_{ij} = \frac{1}{2} \bigg(\frac{\partial {v_s}_i}{\partial x_j} - \frac{\partial {v_s}_j}{\partial x_i} \bigg)$\\
			Granular Phase Plastic Flow Rate & $\displaystyle \tilde{D}^p_{ij} = \frac{\dot{\bar{\gamma}}^p}{2\bar{\tau}} (\tilde{\sigma}_{ij} + \tilde{p}\delta_{ij}) + \tfrac{1}{3}(\beta \dot{\bar{\gamma}}^p + \dot{\xi}_1 + \dot{\xi}_2)\delta_{ij}$\\
			Equivalent Granular Shear Stress & $\bar{\tau} = \sqrt{\tfrac{1}{2}(\sigma_{kl} + \tilde{p}\delta_{kl})(\sigma_{kl} + \tilde{p}\delta_{kl})}$\\
			Granular Pressure & $\tilde{p} = -\tfrac{1}{3} \tilde{\sigma}_{kk}$\\
			Dilation Angle &
			$\beta = K_3\, (\phi-\phi_{eq})$\\
			Critical State Packing Fraction & $\displaystyle \phi_{eq} = \frac{\phi_m}{1+aI_m}$\\
			Frictional model for $\phi_m$ & $\displaystyle \phi_m = \phi_j + (\phi_c - \phi_j)f$\\
			Frictional model for $a$ & $\displaystyle a = a_0 + (a_\infty - a_0)f$\\
			Evolution of Granular Structure & $\displaystyle \dot{f} = K_0 \dot{\bar{\gamma}}^p \bigg( \frac{\bar{\tau}}{\tau^*} \bigg)^{3/2}(f_m - f) - K_0 \bigg( \dot{\bar{\gamma}}^p + \frac{\bar{\tau}}{\eta_B} + \dot{\xi}_\epsilon \bigg)f$\\
			Internal Friction Coefficient & $\displaystyle \mu_p = \mu_1 + \frac{\mu_2 - \mu_1}{1 + (b/I_m)} + \tfrac{5}{2} \bigg(\frac{\phi I_v}{aI_m}\bigg)$\\
			Inertial Numbers & $\displaystyle I_v = \frac{\eta_0 \dot{\bar{\gamma}}^p}{\tilde{p}}, \quad I = \dot{\bar{\gamma}}^p d \sqrt{\frac{\rho_s}{p}}, \quad I_m = \sqrt{I^2 + 2I_v}$\\[15pt]
			Granular Shear Flow Rule & $\displaystyle
			\begin{aligned}
			&f_1 = \bar{\tau} - \max\big((\mu_p + \beta)\tilde{p},\ 0\big)\\[1pt]
			&f_1 \leq 0, \qquad \dot{\bar{\gamma}}^p \geq 0, \qquad f_1 \dot{\bar{\gamma}}^p = 0
			\end{aligned}$\\[15pt]
			Granular Separation Rule & $\displaystyle
			\begin{aligned}
			&f_2 = -\tilde{p}\\[1pt]
			&f_2 \leq 0, \qquad \dot{\xi}_1 \geq 0, \qquad f_2 \dot{\xi}_1 = 0
			\end{aligned}$\\[15pt]
			Granular Compaction Rule & $\displaystyle
			\begin{aligned}
			&f_3 = g(\phi) \tilde{p} - (a\phi)^2 \big[ \zeta^2 d^2 \rho_s + 2 \eta_0 \zeta \big]\\[1pt]
			&f_3 \leq 0, \qquad \dot{\xi}_2 \leq 0, \qquad f_3 \dot{\xi}_2 = 0\\[3pt]
			&\zeta = \dot{\bar{\gamma}}^p - K_4 \dot{\xi}_2\\
			&g(\phi) = \bigg\{\def\arraystretch{1}\begin{matrix}
			(\phi_m-\phi)^2 & \mathrm{if} \quad \phi < \phi_m\\
			0 & \mathrm{if} \quad \phi \geq \phi_m
			\end{matrix}
			\end{aligned}$\\
			\bottomrule
		\end{tabular}
	\end{center}
\end{table}

\subsection*{Determining $H$ from \cite{mari2014}}
The proposed model for the evolution of granular micro-structure $f$ in dense suspensions has the following form,
\begin{equation}
\frac{\dot{f}}{K_0 \dot{\gamma}} = H(f_m - f) - Sf
\end{equation}
with H a hardening parameter, S a softening parameter, $f_m$ a limiting value for $f$ (taken here to be 1), and $K_0$ a scaling term. This expression for $\dot{f}$ has a zero which corresponds to steady behavior of the granular mixture,
\begin{equation} \label{eqn:steadyf}
\dot{f}=0 \quad \implies \quad f = \frac{H f_m}{H+S}.
\end{equation}

In this work, we have posited that
\begin{equation}
H = \hat{H}(\bar{\tau}/\tau^*),
\end{equation}
with $\tau^*$ a repulsive stress scale and $\bar{\tau}$ a measure of the applied granular stress such that the rate of hardening of the granular micro-structure depends on the magnitude of the applied granular stress overcoming inter-granular repulsion. Additionally, we have posited that the softening behavior can be described as follows,
\begin{equation}
S = \bigg( 1 + \frac{\dot{\xi}_B}{\dot{\gamma}} + \frac{\dot{\xi}_\epsilon}{\dot{\gamma}} \bigg),
\end{equation}
with a constant term for the rate of structural decay in shear, $\dot{\xi}_B$ the rate of structural breakdown due to buckling of force chains, and $\dot{\xi}_\epsilon$ the rate of structural breakdown due to diffusion.

\begin{figure}[H]
	\centering
	\includegraphics[scale=1.0]{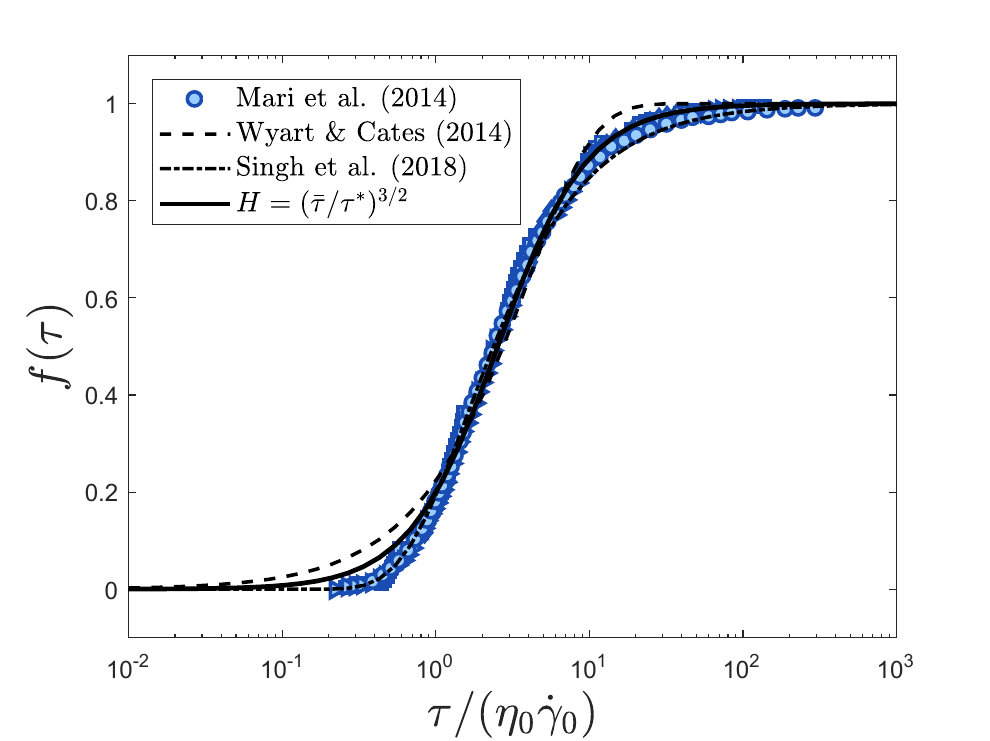}
	\caption{Comparison of simulated values of $f$ in \cite{mari2014} (blue markers), the model for $f(\tau)$ in \cite{wyart2014} (dashed line), the model for $f(\tau)$ in \cite{singh2018} (dashed/dotted line), and the steady model proposed in this work for $H = (\bar{\tau}/\tau^*)^{3/2}$ (solid line).}
	\label{fig:mari}
\end{figure}

If $\dot{\xi}_B$ and  $\dot{\xi}_\epsilon$ are taken to be negligible in comparison to the shearing rate $\dot{\gamma}$, then we find that the steady behavior of $f$ is a function of stress only,
\begin{equation}
\dot{\xi}_B = \dot{\xi}_\epsilon = 0, \quad \dot{f} = 0 \quad \implies \quad f = f(\bar{\tau}) = \frac{\hat{H}(\bar{\tau}/\tau^*)}{\hat{H}(\bar{\tau}/\tau^*) + 1}.
\end{equation}
In dense suspensions, the granular stress $\bar{\tau}$ will account for most of the total mixture stress $\tau$ such that $f(\bar{\tau}) \approx f(\tau)$. This direct dependence of $f$ on the mixture stress $\tau$ is in agreement with the models proposed in \cite{wyart2014} ($f(\tau) \approx 1 - e^{-\tau/\tau^*}$) and \cite{singh2018} ($f(\tau) \approx e^{-\tau^*/\tau}$) as well as the simulated flows reported in \cite{mari2014}. As shown in figure \ref{fig:mari}, there is strong agreement between these models and our prediction for $f(\bar{\tau})$ when
\begin{equation}
H = \hat{H}(\bar{\tau}/\tau^*) = \bigg(\frac{\bar{\tau}}{\tau^*} \bigg)^{3/2}.
\end{equation}

\subsection*{Estimating $\phi$ from $\phi_W$ in \cite{hermes2016}}
Due to the significant swelling of starch particles suspended in fluid, the mixture mass fractions $\phi_W$ is reported for the suspensions studied in \cite{hermes2016} instead of the true volume fractions $\phi$ with,
\begin{equation}
\phi_W = \frac{m_{cs}}{m_{cs} + m_{l}},
\end{equation}
and $m_{cs}$ the mass of cornstarch in the mixture and $m_l$ the mass of suspending fluid in the mixture. To convert this to solid volume fraction $\phi$ for model fitting, we use the method described in \cite{peters2016} that attempts to account for pore space and particle swelling as follows,
\begin{equation}
\phi =  \frac{(1 + \lambda) (1-\beta)(m_{cs}/\rho_{cs})}{(1-\beta)(m_{cs}/\rho_{cs}) + (m_l/\rho_l) + \beta (m_{cs}/\rho_w)},
\end{equation}
with $\lambda \approx 0.3$, $\beta \approx 0.1$, $\rho_{cs}$ the density of a cornstarch grain, $\rho_l$ the density of the suspending fluid, and $\rho_w$ the density of water. This expression can be written equivalently as follows,
\begin{equation}
\phi = \frac{(1 + \lambda) (1-\beta)}{(1-\beta) + (\frac{1-\phi_W}{\phi_W})(\rho_{cs}/\rho_l) + \beta (\rho_{cs}/\rho_w)},
\end{equation}
and is used to determine the values of $\phi$ shown in this work.

\subsection*{Fitting $a_0$ and $\phi_j$ to Steady Shearing Data}
\begin{figure}[H]
	\centering
	\includegraphics[scale=0.6]{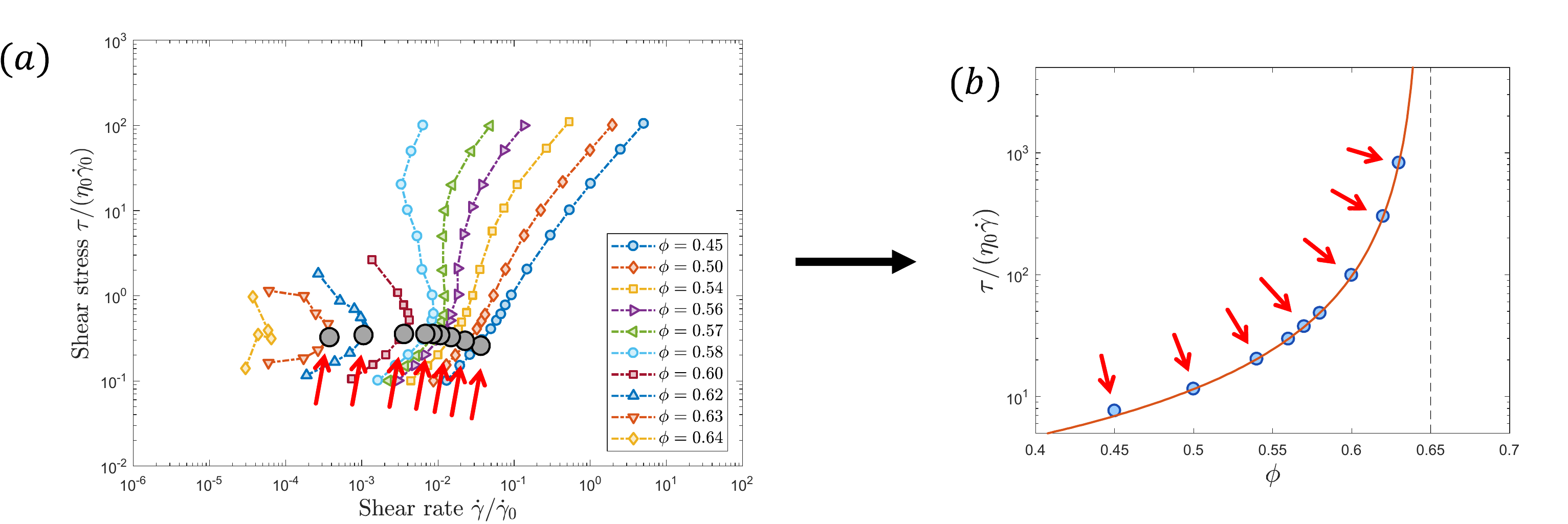}
	\caption{(a) Steady shearing behavior of $\mu = 1$ simulations from \cite{singh2018} colored by volume fraction $\phi$ (see legend). Black circular markers represent minimum viscosity measured in mixture at each simulated volume fraction. (b) The blue circles represent the minimum viscosity measurements found in (a), the red line represents the best fit of \eqref{eqn:phi_j} to this data, and the dashed black line represents the asymptote associated with the value of $\phi_j$.}
	\label{fig:phi_j}
\end{figure}

Here we describe a method for determining $a_0$ and $\phi_j$ from steady shearing data. Recall the expression for the effective viscosity of a mixture in steady shearing flow from \cite{baumgarten2018},
\begin{equation}
\frac{\tau}{\eta_0 \dot{\gamma}} = \eta_r =  1 + \frac{5}{2} \phi \bigg(\frac{\phi_m}{\phi_m - \phi}\bigg) + 2 \mu_c \bigg(\frac{a\phi}{\phi_m - \phi}\bigg)^2,
\end{equation}
and the expressions for $a$ and $\phi_m$ from \cite{singh2018},
\begin{equation}
\phi_m = \hat{\phi}_m(f) = \phi_j + (\phi_c - \phi_j)f,
\end{equation}
\begin{equation}
a = \hat{a}(f) = a_0 + (a_\infty - a_0)f.
\end{equation}
If we consider the behavior of a mixture at relatively low shearing rates and stresses, we expect that $f \to 0$. In this limit, we find an expression for effective viscosity $\eta_r$ that is identical to that from \cite{boyer2011} with $\phi_m = \phi_j$ and $a = a_0$,
\begin{equation} \label{eqn:phi_j}
f \to 0 \quad \implies \quad \eta_r(\phi) =  1 + \frac{5}{2} \phi \bigg(\frac{\phi_j}{\phi_j - \phi}\bigg) + 2 \mu_c \bigg(\frac{a_0 \phi}{\phi_j - \phi}\bigg)^2.
\end{equation}
We can fit this expression to the low viscosity measurements pulled from experiments (see figure \ref{fig:phi_j}) to determine the best values of $a_0$ and $\phi_j$. (Note that $\mu_c$ is chosen to reflect the dry granular behavior of the granular particles and is not fit using these curves.)

\subsection*{Fitting $a_\infty$ and $\phi_c$ to Steady Shearing Data}
\begin{figure}[H]
	\centering
	\includegraphics[scale=0.6]{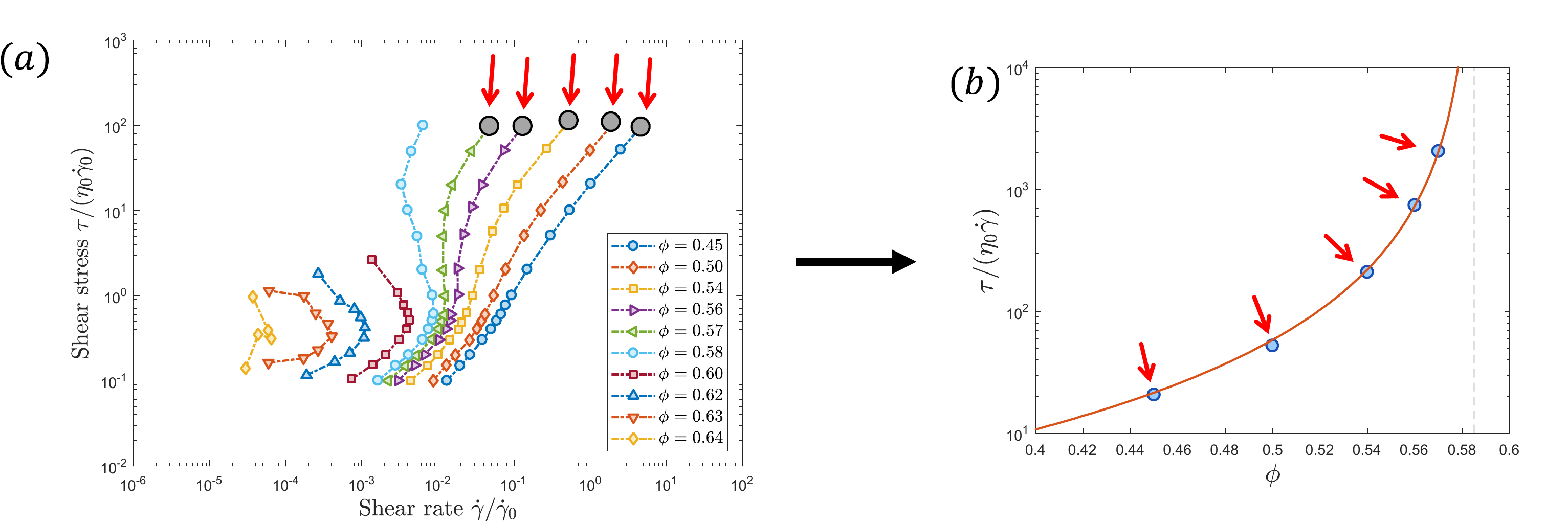}
	\caption{(a) Steady shearing behavior of $\mu = 1$ simulations from \cite{singh2018} colored by volume fraction $\phi$ (see legend). Black circular markers represent maximum viscosity measured in mixture at several simulated volume fractions. (b) The blue circles represent the maximum viscosity measurements found in (a), the red line represents the best fit of \eqref{eqn:phi_c} to this data, and the dashed black line represents the asymptote associated with the value of $\phi_c$.}
	\label{fig:phi_c}
\end{figure}

As in the previous section, we can fit $a_\infty$ and $\phi_j$ to experimental data if we consider the behavior of a mixture at relatively high shearing rates and stresses where we expect that $f \to 1$. In this limit, we find an expression for effective viscosity $\eta_r$ that is identical to that from \cite{boyer2011} with $\phi_m = \phi_c$ and $a = a_\infty$,
\begin{equation} \label{eqn:phi_c}
f \to 1 \quad \implies \quad \eta_r(\phi) =  1 + \frac{5}{2} \phi \bigg(\frac{\phi_c}{\phi_c - \phi}\bigg) + 2 \mu_c \bigg(\frac{a_\infty \phi}{\phi_c - \phi}\bigg)^2.
\end{equation}
We can fit this expression to the high viscosity measurements pulled from experiments (see figure \ref{fig:phi_c}) to determine the best values of $a_\infty$ and $\phi_c$. (Note that we limit the range of volume fractions considered to flow curves which have a well defined secondary plateau.)

\subsection*{Fitting $\dot{\gamma}_{\text{DST}}$ to Steady Shearing Data}
\begin{figure}[H]
	\centering
	\includegraphics[scale=0.6]{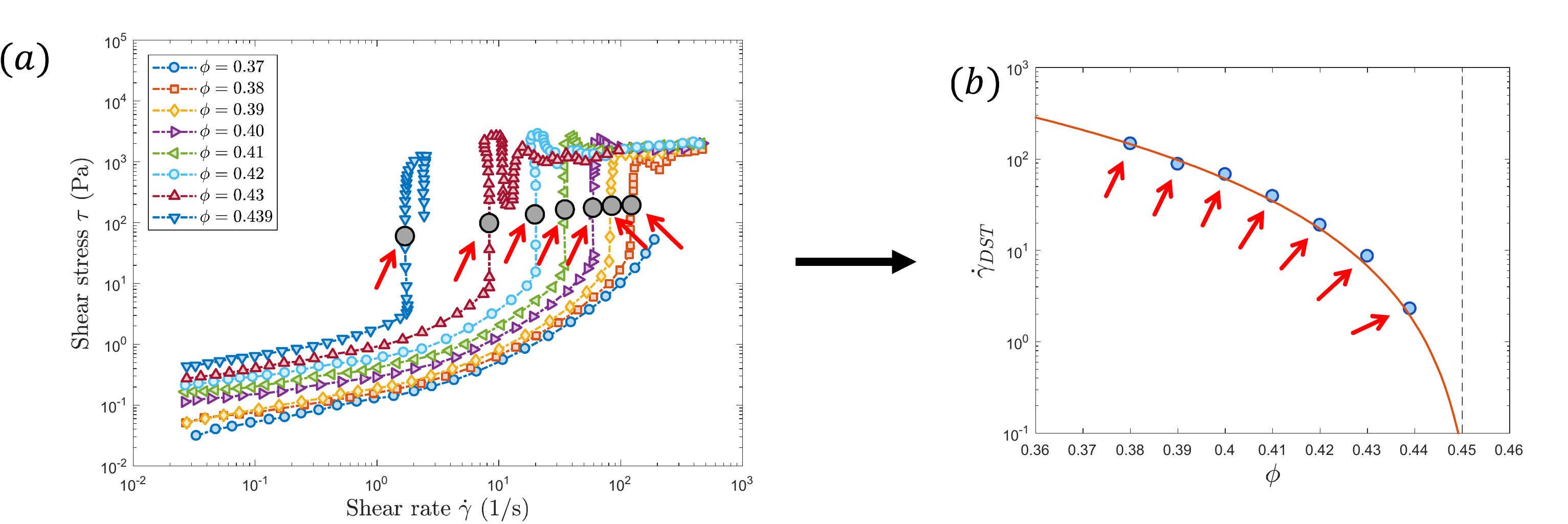}
	\caption{(a) Steady shearing behavior of cornstarch mixtures as reported in \cite{fall2015} colored by volume fraction $\phi$ (see legend). Black circular markers show DST region and associated shearing rates. (Note that the upper plateau of these curves is the result of wall slip.) (b) The blue circles represent DST shearing rate measurements found in (a), the red line represents the best fit of \eqref{eqn:dst} to this data, and the dashed black line represents the asymptote associated with the value of $\phi_j$.}
	\label{fig:gdst}
\end{figure}

For rheologically stable flows, the parameters $\tau^*$ and $\eta_B$ are fit by observation with $\eta_B$ generally taken to be very large (see fits for \cite{singh2018} in main document). For rheologically chaotic flows, large applied shear stresses at packing fractions above $\phi_c$ appear to cause unbounded DST (mixture breakdown and wall slip limit the range of stresses that can be experimentally measured). To fit $\eta_B$ to such flows, we begin by recalling the proposed form of $\hat{f}_m(\phi)$ for volume fractions in the range of $\phi_c \leq \phi \leq \phi^*$,
\begin{equation}
\hat{f}_m(\phi) = \frac{\phi_j - \phi}{\phi_j - \phi_c}, \text{\quad if \quad} \phi_c \leq \phi \leq \phi^*.
\end{equation}
Combining this expression with \eqref{eqn:steadyf}, we find,
\begin{equation}
\dot{f}=0 \quad \implies \quad f = \frac{H}{H+S} \frac{\phi_j - \phi}{\phi_j - \phi_c}, \text{\quad if \quad} \phi_c \leq \phi \leq \phi^*,
\end{equation}
and together with the expression for $\phi_m$ from \cite{singh2018},
\begin{equation}
\phi_m = \phi_j + (\phi - \phi_j)\frac{H}{H+S}, \text{\quad if \quad} \phi_c \leq \phi \leq \phi^*.
\end{equation}
Substituting this expression into the an expression for the effective granular viscosity in \cite{baumgarten2018}, we have,
\begin{equation}
\frac{\bar{\tau}}{\eta_0 \dot{\gamma}} = \frac{5}{2} \bigg( \frac{\phi^2}{\phi_j - \phi} \bigg) \bigg(\frac{H+S}{S}\bigg) + 2 \mu_c \bigg( \frac{a \phi}{\phi_j - \phi} \bigg)^2 \bigg(\frac{H+S}{S}\bigg)^{2}, \text{\quad if \quad} \phi_c \leq \phi \leq \phi^*.
\end{equation}
We continue by recalling the forms of $H$ and $S$ proposed in this work; from these, it can be shown that,
\begin{equation}
\lim_{\bar{\tau} \to \infty}{H} = \bigg(\frac{\bar{\tau}}{\tau^*}\bigg)^{3/2},
\end{equation}
and,
\begin{equation}
\lim_{\bar{\tau} \to \infty}{S} = \frac{\bar{\tau}}{\eta_B \dot{\gamma}}.
\end{equation}
All of this combines together to give the following result,
\begin{equation}
\bar{\tau} \to \infty \quad \implies \quad \frac{\bar{\tau}}{\eta_0 \dot{\gamma}} = 2 \mu_c \bigg(\frac{\hat{a}(f_m) \phi}{\phi_j - \phi}\bigg)^2 \frac{\bar{\tau} (\eta_B \dot{\gamma})^2}{(\tau^*)^3}, \text{\quad if \quad} \phi_c \leq \phi \leq \phi^*,
\end{equation}
which can be reduced to an equation of $\dot{\gamma}$ and $\phi$ only to yield the expression for $\dot{\gamma}_{\text{DST}}$ shown in the main document,
\begin{equation} \label{eqn:dst}
\lim\limits_{\bar{\tau} \to \infty} \dot{\gamma} = \dot{\gamma}_{\text{DST}} = \frac{\tau^*}{\eta_0} \bigg(2 \mu_c \bigg(\frac{\eta_B}{\eta_0} \frac{\hat{a}(f_m) \phi}{(\phi_j - \phi)}\bigg)^{2} \bigg)^{-\tfrac{1}{3}}, \text{\quad if \quad} \phi_c \leq \phi \leq \phi^*.
\end{equation}
We can then fit this expression to the experimentally measured shearing rates associated with observed DST to determine a reasonable form for $\eta_B$ (see figure \ref{fig:gdst}). Observation of the trends of $\dot{\gamma}_{\text{DST}}$ in the literature (and $\dot{\gamma}_c$ in \cite{barnes1989}) suggest the form proposed for $\eta_B$ in the main document,
\begin{equation}
\eta_B = \hat{\eta}_B(\phi) = \bigg(\sum_{i=1}^{i_{\text{max}}} \varphi_i (\phi_j - \phi)^{\alpha_i} \bigg)^{-1},
\end{equation}
such that the resistance to buckling induced degradation, $\eta_B$, grows as the volume fraction $\phi$ increases toward $\phi_j$ at a rate controlled by $\varphi_i$ and $\alpha_i$.

{
	
	\subsection*{Ringing Instability in Cornstarch-Water Simulations using MPM}
	The material point method (MPM; see \cite{sulsky1994}) is a numerical scheme for solving dynamic mechanics problems and excels when history dependent materials undergo large deformations. Unlike standard Lagrangian methods, a static background grid is used to represent the material velocities and weak form test functions; using a fixed grid avoids the problems of mesh distortion that often accompany large material deformations. Unlike standard Eulerian methods, a fixed set of material point tracers are used to represent history dependent properties (e.g.\ $\tilde{\sigma}_{ij}$, $f$, $\phi$) and act as quadrature points for weak form integration; tracking history dependent quantities at fixed points in the material avoids the errors associated with numerical advection schemes. These two features of MPM lead to an unfortunate side effect which is common to many `particle methods': the ringing instability (see \cite{gritton2014}).
	
	The ringing instability is an accumulated error in the material state represented on the material point tracers. This error is allowed to accumulate in the null space of the material-point-to-grid mapping matrix. At each time-step, the material state as represented on the material point tracers must be `mapped' or integrated in order to update the grid representation of velocity. In order to maintain high accuracy when integrating the weak form equations, we often require that the number of points $N_p$ is greater than the number of grid coefficients $N_n$ (often equal to the number of nodes); this guarantees the existence of a null space in which error can grow. In order to minimize the effect of this error on the quality of the results presented in this work, we have generally chosen elastic moduli $G$ and $K$ (and the fluid bulk modulus $\kappa$) which are `large enough' to avoid significant elastic deformations, but not `so large' as to produce significantly spurious stress fields (see figure \ref{fig:stress_error} where example pressure fields, $\tilde{p}$, beneath the elastic wheel in the `running on oobleck' simulations are shown for three different sets of elastic moduli).
	
	\begin{figure}[H]
		\centering
		\includegraphics[scale=0.45]{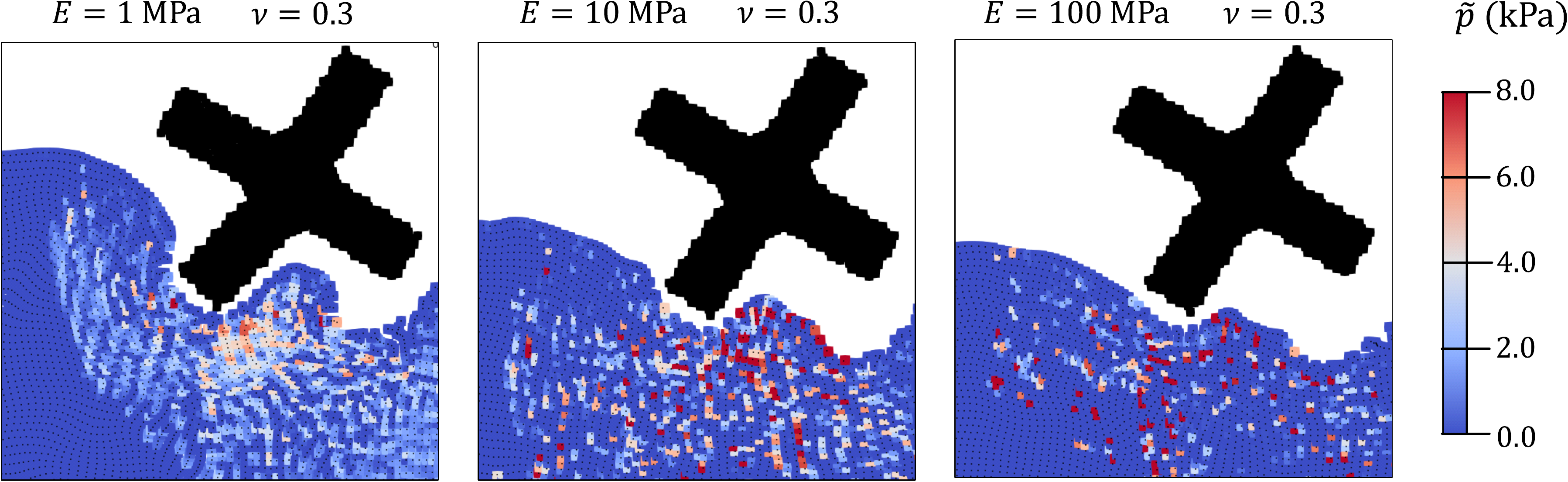}
		\caption{Pressure fields below the elastic wheel in the `running on oobleck' simulations described in the main document with three different elastic moduli (with $K = E/(3(1-2\nu))$ and $G = E/(2(1+\nu))$). Snapshots are taken at $t = 1$s. In all simulations, the elastic deformation within the mixture remains below 1\%. As the elastic modulus of the granular phase is increased, the quality of the spatial stress field degrades. The likely reason for this breakdown is coupling between the ringing instability in MPM with the highly non-linear material model proposed in this work.}
		\label{fig:stress_error}
	\end{figure}
	
	\section*{A Method for Determining $G$ and $K_0$}
	In the model presented in this work, we have assumed that the material which constitutes the granular phase of the mixture is \textit{elastically stiff}. This assumption follows from two observations: (i) the individual grains in mixtures which exhibit DST have elastic moduli on the order of GPa (e.g.\ poly(methyl methacrylate)) and (ii) the largest stresses measured in these mixtures are often smaller than a few MPa. If the elastic moduli $G$ and $K$ of the bulk granular material are of a similar order to the elastic moduli of the individual grains, then it is reasonable to expect that $G$ and $K \gg \bar{\tau}$. For this reason, we have neglected to find realistic values of $G$ and $K$ for use in our model and have assumed that most of the observed dynamics of these mixtures are dominated by plastic flow ($\dot{\bar{\gamma}}^p$) instead of elastic deformation.
	
	In this section, we describe the behavior of our proposed model as it relates to stress relaxation in the mixture after the cessation of simple, quasi-two-dimensional shearing flow for the purposes of experimentally determining the values of $G$ and $K_0$. For reference, the basic equation of our model can be found in table \ref{tab:governing_equations}. If the granular material is sheared to steady-state, then it can be shown that $\phi < \phi_m$, $\dot{f} = 0$, and,
	\begin{equation} \label{eqn:gdp_est}
	\dot{\bar{\gamma}}^p = \frac{\bar{\tau}}{\eta_0} \bigg[2\mu_1 \frac{(a\phi)^2}{(\phi_m - \phi)^2} + \frac{5}{2} \frac{\phi}{\phi_m - \phi}\bigg]^{-1}.
	\end{equation}
	Given a steady shear stress measurement of $\bar{\tau_0}$ (and assuming $\phi_c$, $\phi_j$, $a_0$, $a_\infty$, $\tau^*$, and $\eta_B$ have already been determined), it is then possible to find the steady value of $f$, $f_0$.
	
	With the steady response of the system known, we can then consider its behavior when the material is stopped suddenly; that is, we are interested in the time-accurate measurement of the shear stress, $\bar{\tau}(t)$, after the applied shearing rate is set to zero, $D_{ij} = W_{ij} = 0$. In this regime, we define $\dot{\bar{\gamma}}^p$ according to \eqref{eqn:gdp_est}, $\dot{f}$ according to the rules in table \ref{tab:governing_equations}, and $\dot{\bar{\tau}}$ as follows (derived from expressions in table \ref{tab:governing_equations}):
	\begin{equation} \label{eqn:tau_dot}
	\dot{\bar{\tau}} = - G \dot{\bar{\gamma}}^p.
	\end{equation}
	We can then integrate \eqref{eqn:tau_dot} and the expression for $\dot{f}$ in table \ref{tab:governing_equations} to find,
	\begin{equation}
	\begin{aligned}
	\bar{\tau}(t) &= \bar{\tau}_0 + \int_0^t \dot{\bar{\tau}} dt,\\
	f(t) &= f_0 + \int_0^t \dot{f} dt.
	\end{aligned}
	\end{equation}
	
	Example stress relaxation curves for $\bar{\tau}(t)$ at varying values of $G$ and $K_0$ can be found in figure \ref{fig:decay}. Prior work on examining the relaxation characteristics of these mixtures can be found in \cite{dheane1993} and \cite{maharjan2017}. An important note about the results found in those works is that the characteristic stress relaxation time is on the order of tens of seconds for some of the mixtures considered. In order for our model to reproduce those relaxation times (see figure \ref{fig:decay}(a)), the granular shear modulus $G$ needs to be on the order of 10 kPa, much smaller than what we would normally expect. This time-scale difference suggests that the elastic behavior of these mixtures is more complicated than simple linear elasticity and further work on our model will be necessary to capture this behavior. However, the qualitative similarity between the curves shown in figure \ref{fig:decay} and those found in \cite{maharjan2017} is a promising indication that we are capturing the correct phenomena.
	
	\begin{figure}[H]
		\centering
		\includegraphics[scale=0.4]{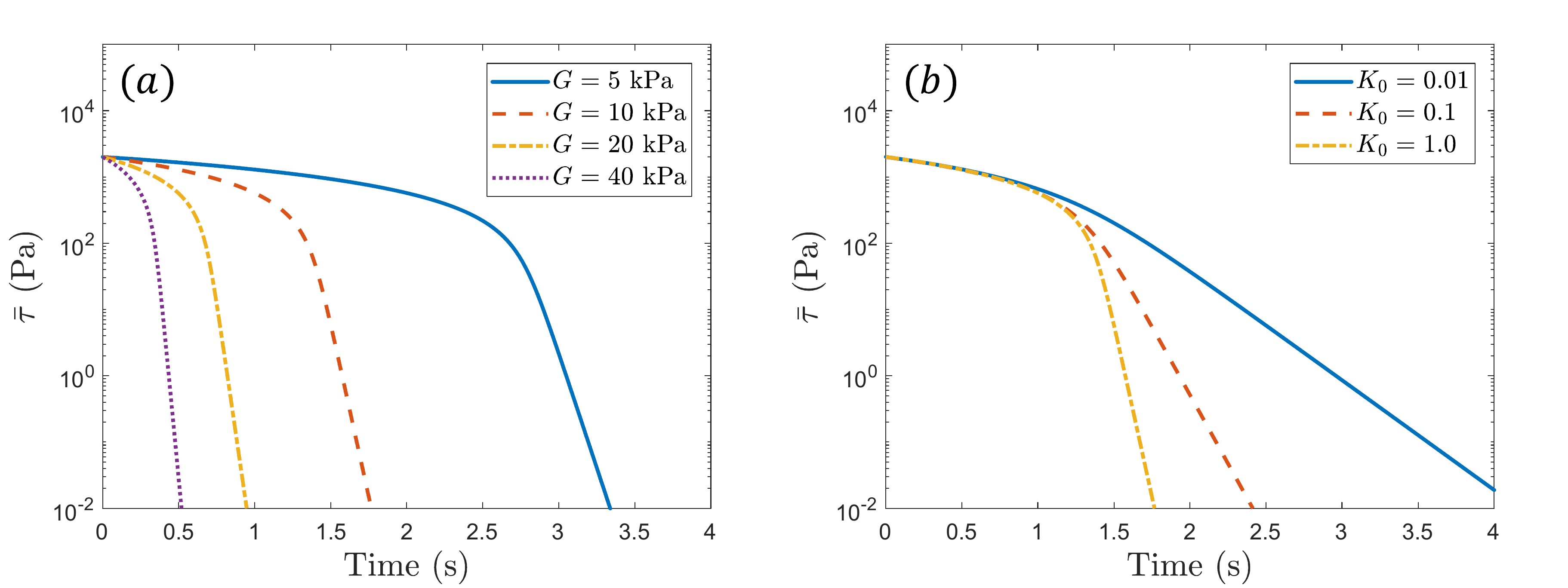}
		\caption{Example stress relaxation curves after the cessation of flow. All curves are generated using the steady-state material parameters determined in the main document for the experimental results found in \cite{hermes2016} at a packing fraction of $\phi = 0.52$. (a) Example curves at varying granular shear moduli $G$. (b) Example curves at varying $K_0$.}
		\label{fig:decay}
	\end{figure}
}

\clearpage

\end{document}